\documentclass[11pt]{article}
\pdfoutput=1

\usepackage{amsmath}
\usepackage{amssymb}
\usepackage{amsfonts}
\usepackage{mathrsfs}

\usepackage{mathrsfs}
\usepackage{fullpage}
\usepackage{setspace}
\usepackage{bm}
\usepackage{bbm}
\usepackage{relsize}
\usepackage{adjustbox}
\usepackage{multirow}
\usepackage{cite}
\usepackage{filecontents}
\usepackage{scalerel}

\usepackage[normalem]{ulem}
\usepackage{enumerate}
\usepackage{yfonts}
\usepackage{psfrag}
\usepackage{array}
\usepackage{booktabs}
\newcolumntype{N}{>{\centering\arraybackslash}m{.5in}}
\newcolumntype{G}{>{\centering\arraybackslash}m{2in}}

\usepackage[latin1]{inputenc}
\usepackage{graphicx}
\usepackage{cancel}
\usepackage{slashed}
\usepackage{mathtools}

\usepackage[font=small,labelfont=bf]{caption}
\usepackage{subcaption}

\usepackage[colorlinks=true]{hyperref}
\hypersetup{
    unicode=false,          
    pdftoolbar=true,        
    pdfmenubar=true,        
    pdffitwindow=false,     
    pdfstartview={FitH},    
    pdftitle={
    Imprints of extra dimensions in eccentric extreme mass-ratio inspiral gravitational waveforms}
    pdfauthor={Shailesh Kumar, Tieguang Zi and Arpan Bhattacharyya},     
    pdfnewwindow=true,      
    colorlinks=true,       
    linkcolor=blue,          
    citecolor=red,        
    filecolor=magenta,      
    urlcolor=blue           
    }

\usepackage{hyperref}
\hypersetup{colorlinks=true}

\def\equationautorefname~#1\null{%
	Eq.~(#1)\null
}
\def\figureautorefname~#1\null{%
	Fig.~#1\null
}
\def\tableautorefname~#1\null{%
	Table.~#1\null
}
\def\sectionautorefname~#1\null{%
	Section #1\null
}
\def\appendixautorefname~#1\null{%
	Appendix #1\null
}

\begin{document}

\numberwithin{equation}{section}
{
\begin{titlepage}
\begin{center}

\hfill \\
\hfill \\
\vskip 0.75in

{\Large {\bf
Imprints of extra dimensions in eccentric extreme-mass-ratio inspirals  gravitational waveforms
}}

\vskip 0.2in

{\large Shailesh Kumar${}$\footnote{\href{mailto: shailesh.k@iitgn.ac.in}{shailesh.k@iitgn.ac.in}}$^{a}$,
Tieguang Zi${}$\footnote{\href{mailto: zitg@scut.edu.cn}{zitg@scut.edu.cn, corresponding author}}}$^{b,c}$,
Arpan Bhattacharyya${}$\footnote{\href{mailto:  abhattacharyya@iitgn.ac.in}{abhattacharyya@iitgn.ac.in}}$^{a}$

\vskip 0.2in

{\it ${}$$^a$Indian Institute of Technology, Gandhinagar, Gujarat-382355, India\\ ${}$
\it ${}$$^b$ Department of physics, Nanchang University,
Nanchang, 330031, China\\ ${}$
\it ${}$$^c$ Center for Relativistic Astrophysics and High Energy Physics, Nanchang University, Nanchang, 330031, China}

\vskip.5mm

\end{center}

\vskip 0.2in

\begin{center}
{\bf ABSTRACT }
\end{center}
Studies regarding extra-dimensions have been of great interest in modern theoretical physics, including their observational consequences from future gravitational wave (GW) observatories. In this direction, extreme-mass-ratio inspirals (EMRIs), attracting considerable interest in GW astronomy and fundamental physics, can potentially provide a useful platform for the search of extra dimensions. In this paper, we examine a rotating braneworld black hole in the context of equatorial eccentric EMRIs and attempt to provide an order of magnitude analysis for the extra-dimensional parameter termed `tidal charge'. We estimate GW fluxes for the dominant mode and determine the impact of the tidal charge parameter on the orbital evolution. We further evaluate the prospects of detecting such a parameter through mismatch computation.
We observe a significant enhancement in the mismatch as the value of orbital eccentricity or tidal charge parameter increases; the phenomenon becomes more obvious for rapidly rotating massive black holes. Thus, the study suggests that eccentric EMRIs can potentially probe the existence of extra dimensions with future low-frequency detectors such as the Laser Interferometer Space Antenna (LISA).
\vfill


\end{titlepage}
}

\newpage

\section{Introduction}
The pioneering discovery by the Laser Interferometer Gravitational Wave Observatory (LIGO), the direct detection of gravitational waves (GWs) triggered by a binary black hole (BBH) system, has paved the way for key benchmarks in the field of astrophysics, bringing foundational impacts and developments in GW astronomy \cite{PhysRevLett.116.061102}. There have been plenty of such cataclysmic events detected to date, which have advanced our domain of interest, knowledge, and insights \cite{PhysRevD.100.062006, LIGOScientific:2016lio, LIGOScientific:2020tif, LIGOScientific:2021sio, LIGOScientific:2019fpa, LIGOScientific:2016sjg, LIGOScientific:2017bnn}. The observational consequences of GWs allow us for detailed testing of General Relativity (GR) in extreme environments and also enable us to enhance and understand the formation as well as properties of various black hole populations \cite{LIGOScientific:2018jsj}. These waves carry astrophysical characteristics of BBH systems, such as their masses and spins, facilitating us to constrain the parameters with existing and future GW detectors. Out of three stages of a binary system\textemdash inspiral, merger, and ringdown\textemdash the inspiral phase is one of the very crucial phases due to several reasons, namely detecting a strong GW signal, pre-merger information like masses, spins, and orbital parameters, and the test of GR under strong gravity regimes \cite{PhysRevLett.116.061102, Sathyaprakash:2009xs, Bailes:2021tot}. Inspiral phase can be modelled with numerous current techniques, for example, black hole perturbation (BHP) and post-Newtonian (PN) \cite{Pound:2021qin, Blanchet:2013haa},  permitting for more accurate comparisons between estimated theoretical predictions and observed GW signal. Therefore, the present article solely focuses on the inspiral phase of a particular BBH system, termed `extreme-mass-ratio inspirals' (EMRIs).

An EMRIs is a two-body system that consists of a stellar-mass object (mass $\mu$) called secondary that inspirals a massive black hole (MBH with mass $M$) termed primary, whose mass-ratio ($q$) lies in the range of $q\equiv \mu/M \in \{10^{-7},10^{-4}\}$. EMRIs have become one of the most prominent candidates in black hole physics that capture attention from various perspectives due to their unique properties, ranging from the astrophysical environment \cite{PhysRevD.105.L061501, PhysRevLett.129.241103, Destounis:2022obl, Rahman:2023sof, Duque:2024mfw, Dai:2023cft, Miller:2025yyx} to the tests of GR and beyond \cite{Cardenas-Avendano:2024mqp, PhysRevD.102.064041, PhysRevLett.126.141102, Kumar:2024utz, Zhang:2024csc, Kumar:2024our, Zi:2023qfk, Zi:2024jla, Kumar:2024dql, Zi:2022hcc, Zi:2023pvl, Fu:2024cfk, Qiao:2024gfb, Babichev:2024hjf}. These systems not only advance our understanding of the strong or weak gravity regimes of black holes but also offer us the ability to predict the observables with high accuracy. As these sources are very suitable for low-frequency GW detectors, the recent progress indicates the potential for their detectability using space-based detectors like the Laser Interferometer Space Antenna (LISA) \cite{amaroseoane2017laserinterferometerspaceantenna} and TianQin \cite{TianQin:2015yph, TianQin:2020hid}. As stated before, the secondary object exhibits the inspiralling motion in the background of the primary; one can treat such a system in a perturbative manner through BHP or PN techniques. The present article adopts the Teukolsky framework \cite{1973ApJ, 1972ApJ, PhysRevLett.29.1114}, a BHP approach that has a major advantage of being a scalar equation and very suitable for probing extreme environments of gravity as well as serving the purpose of mathematical simplifications. 

On the other hand, exploring spatial geometries/dimensions beyond the familiar three has gained substantial attraction and become a central focus of modern theoretical physics, mainly driven by string theory and M-theory frameworks, which provide a robust and unified approach to integrating gravity within a higher-dimensional spacetime \cite{Shiromizu:1999wj, Rubakov:2001kp}. The idea of extra spatial dimensions, nearly as old as general relativity (GR), was first explored by Kaluza and Klein in the 1920s to unify gravity and electromagnetism \cite{Kaluza:1921tu, Klein1926} and was further motivated by string theory, suggesting that our universe may have more than three dimensions. Recent trends suggest that superstring theory can argue the existence of the universe within an eleven-dimensional structure, where dualities link the distinct string theories together \cite{Shiromizu:1999wj, Antoniadis:1998ig}. Several notable studies have originated from this foundation, from exploring its implications to profoundly impacting our cosmos \cite{Clifton:2011jh, DeFelice:2010aj, Joyce:2014kja}. The understanding of the aforementioned scenario can be made more comprehensible in a five-dimensional picture, where matter fields reside in a four-dimensional geometry, and gravitational forces act in all five dimensions. In this direction, the braneworld concept has become a pivotal idea in modern physics, proposing that our universe could be a multidimensional surface embedded in a higher-dimensional space, offering novel perspectives into gravitational physics \cite{Rubakov:2001kp, Randall:1999ee, Randall:1999vf, Maartens2004}. In other words, the universe is described as a four-dimensional surface, called a `brane,' implanted within a higher-dimensional space known as the `bulk,' exhibiting $Z_{2}$ symmetry. While all matter fields are restricted to the brane, gravity is a unique entity that can propagate into extra dimensions of the bulk. In models such as Randall-Sundrum, extra dimensions can be large or even infinite, where the four-dimensional universe is described as a brane in a higher-dimensional anti-de Sitter spacetime with matter confined to the brane and gravity extending into bulk, further offering potential ways to address issues like the hierarchy problem \cite{Arkani-Hamed:1998jmv, Antoniadis:1998ig, Randall:1999ee, Randall:1999vf, ParticleDataGroup:2018ovx, Dvali:2000hr}.

In the low-energy regime, the effective field equations on the brane mimic the Einstein field equations, encompassing two additional terms: the first term is a local correction appearing from the $Z_{2}$ symmetry of spacetime and the Israel junction conditions, whereas the second is a non-local correction term derived from the bulk Weyl tensor \cite{Shiromizu:1999wj}. Remarkably, the static and spherically symmetric vacuum solution of the effective field equations shares the same mathematical form as the Reissner-Nordstr$\ddot{\textup{o}}$m (RN) black hole solutions in general relativity \cite{Dadhich:2000am, Chamblin:2000ra}. It has also been shown that one can obtain stationary and axisymmetric rotating black hole solutions for a 3-brane within the Randall-Sundrum braneworld framework \cite{Aliev:2005bi}. Such black holes carry a charge termed the `tidal charge' (Q) that distinguishes them from the electric charge of RN and Kerr-Newmann (KN) black holes. Importantly, the tidal charge stems from non-local bulk effects, which can influence the brane geometry even in low-curvature regions. It is understandable that braneworld corrections are relevant at high curvatures, which correspond to smaller black hole masses. However, the situation is more nuanced. In the model, we notice that the tidal charge enters the metric as a correction term $QM^{2}$. This implies that the influence of the correction scales with the black hole mass, indicating that while the curvature falls with increasing black hole mass, the magnitude of the metric correction term increases as $M^{2}$, allowing the tidal charge to still leave a measurable imprint on the SMBH regime. Such an influence can lead to very intricate and promising results with the observation of EMRIs.

Numerous recent literature is available in the context of hunting extra dimensions through distinct astrophysical phenomena \cite{Banerjee:2019nnj, Horvath:2012ru, Zakharov:2018awx, Neves:2020doc, Mishra:2021waw, Vagnozzi:2019apd, Khlopunov:2022jaw, Bohra:2023vls, Zi:2024dpi, Rahman:2022fay}. The prospects of detecting large or infinite extra dimensions in experiments make this concept particularly intriguing and fascinating, where the direct detection of GWs provides a new way to probe extra dimensions through future GW detectors. This motivates us to explore such an aspect in the context of eccentric EMRIs, where an inspiralling object possesses a finite non-zero eccentricity. Also, the inclusion of eccentricity is important as the secondary object in an EMRIs system, with a large initial eccentricity, can significantly affect measurable quantities \cite{Barsanti:2022ana}. Thus, eccentric orbits can make noteworthy contributions to observables and help probe the existence of new degrees of freedom. In this line of endeavour, the present article, focusing on EMRIs, investigates the detection prospects of the tidal charge term with eccentric equatorial orbits of the inspiralling object and examines whether LISA observations can provide more stringent constraints on this parameter, thereby offering insights into extra dimensions.

Let us briefly look at how the article is organized. In Sec. (\ref{BH}), we touch upon the metric details and the eccentric orbital motion of the inspiralling object, as well as numerically obtain the behaviour of the separatrix for distinct values of tidal charge and black hole spin. In Sec. (\ref{perturbation}), we briefly discuss the key equations governing fluxes and orbital evolution within the adiabatic approximation and further discuss the results obtained with different values of tidal charge. Next, Sec. (\ref{wave}) mentions the details of waveform and mismatch for different eccentricities and spin; this provides the stage for detectability of tidal charge parameter with LISA observations, inferring the existence of extra dimensions with the prospects of their detectability. Finally, we conclude our key findings with future aspects in Sec. (\ref{dscn}).


\section{Rotating Braneworld Spacetime} \label{BH}
Braneworld picture provides an interesting framework that exhibits significant deviations from their four-dimensional GR counterparts due to the presence of additional degrees of freedom arising from extra dimensions. In this line of endeavour, the braneworld black holes, described by spherically symmetric or axisymmetric stationary spacetimes, emerge from solving the effective Einstein field equations (EFEs) on the brane \cite{Aliev:2005bi, Dadhich:2000am, Shiromizu:1999wj, Harko:2004ui}. The solution for the spherically symmetric metric on the brane appears similar to the RN case \cite{Dadhich:2000am}, whereas the rotating metric gives rise to a KN-like solution \cite{Aliev:2005bi, Aliev:2009cg}. These solutions capture a charge termed `tidal charge' instead of an electric charge, which contains the imprints of extra dimensions. In this section, we briefly discuss the rotating braneworld black hole and the orbital motion of the inspiralling object in this background.

In order to arrive at the black hole solutions on the brane, as stated before, one can begin with the five-dimensional EFEs that govern the bulk spacetime. When the bulk EFEs are projected onto the brane, a four-dimensional surface, the influence of bulk physics manifests as imprints on the brane. This process accounts for the impact of the extra dimension on the brane without requiring a direct bulk spacetime solution. The effective EFEs on the vacuum brane can be expressed as \cite{Aliev:2005bi, Aliev:2009cg, Harko:2004ui}
\begin{align}\label{effective EFE}
^{(4)}G_{\mu\nu} + E_{\mu\nu} = 0,
\end{align}
where $^{(4)}G_{\mu\nu}$ denotes the Einstein tensor on the brane. $E_{\mu\nu}$ is the electric component of the bulk Weyl tensor, encoding information about the gravitational field beyond the brane. In the case of a static and spherically symmetric spacetime, the exact solution of Eq. (\ref{effective EFE}) corresponds to an RN-like black hole with a tidal charge, as mentioned earlier. When axisymmetry is introduced, a rotating black hole solution is obtained on the brane, with the spacetime geometry expressed as follows \cite{Aliev:2005bi, Harko:2004ui, Zi:2024dpi}:
\begin{align}\label{metric}
ds^{2} = -\frac{\Delta(r)}{\Sigma(r,\theta)} (dt-a\sin^{2}\theta d\phi)^{2}+\frac{\sin^{2}\theta}{\Sigma(r,\theta)}((r^{2}+a^{2})d\phi-a dt)^{2}+\frac{\Sigma(r,\theta)}{\Delta(r)}dr^{2}+\Sigma(r,\theta) d\theta^{2},
\end{align}
where $\Delta(r) = r^{2}-2Mr+a^{2}+QM^{2}$ and $\Sigma(r,\theta) = r^{2}+a^{2}\cos^{2}\theta$. Parameters ($a, M, Q$) are the spin, mass and tidal charge of the black hole. From the metric, it is trivial that if we set $Q=0$, we arrive at the Kerr solution. In general, $Q$ can take positive and negative values. With positive values and replacement $Q\longrightarrow Q^{2}$, the spacetime resembles the KN geometry. We can also see the location of the horizon through $\Delta(r)=0 \Longrightarrow r_{\pm} = M \pm \sqrt{M^{2}-a^{2}-Q}$; where the condition for the presence of the horizon is $M^2 \geq Q+a^2$.

\subsection{Orbital motion}
We notice that the spacetime under consideration (\ref{metric}) exhibits two symmetries that give rise to two constants of motion ($E, L_{z}$), i.e., orbital energy and angular momentum of the point particle moving in the given background. With this, we can derive general expressions for the timelike four geodesic velocities of the particle using the Hamilton-Jacobi equation \cite{Kumar:2024utz}. The related expression can be written as follows
\begin{equation}
\begin{aligned}\label{geodesic}
\mu\Delta\Sigma\frac{dt}{d\tau} =& \Big[\left(a^2+r^2\right) \left(a^2 E-a L_{z}+E r^2\right)+a \Delta \left(L_{z}-a E \sin ^2(\theta )\right) \Big]\,, \\
\mu\Delta\Sigma\frac{d\phi}{d\tau} =& \Big[a \left(a^2 E-a L_{z}+E r^2-E \Delta\right)+L_{z} \csc ^2(\theta ) \Delta\Big]\,, \\
\mu^{2}\Sigma^{2}\Big(\frac{dr}{d\tau}\Big)^{2} =& \Big(E(a^{2}+r^{2})-aL_{z}\Big)^{2}-\Delta\Big(\mu^{2}r^{2}+\mathcal{Q}+(L_z-aE)^2\Big)\,, \\
\mu^{2}\Sigma^{2}\Big(\frac{d\theta}{d\tau}\Big)^{2} =& \Big(\mathcal{Q}+(L_z-aE)^2 - \mu^{2}a^{2}\cos^{2}\theta\Big)-\Big(aE\sin\theta-\frac{L_{z}}{\sin\theta}\Big)^{2}
\end{aligned}
\end{equation}
where $\mu$ is mass of the inspiralling object and $\mathcal{Q}$ is the conventional Carter constant. Note that in subsequent calculations, we will use dimensionless units defined as $\hat{r}=r/M, \hat{E}=E/\mu, \hat{L}_z=L_z/(\mu M), \hat{a}=a/M$ \cite{PhysRevD.102.024041}. However, for convenience, we do not use the `hat' or any specific notation to write distinct quantities; instead, we keep them as it is without the `hat'. One can always re-write the observables in physical units at a later stage. It is apparent from $d\theta/d\tau$ expression in Eq. (\ref{geodesic}) that if we consider equatorial orbits, i.e., $\theta = \pi/2$, the Carter constant turns out to be zero ($d\theta/d\tau=0, \mathcal{Q}=0$) \cite{Glampedakis:2002cb, Glampedakis:2002ya, PhysRevD.61.084004}. Further,
the radial equation of (\ref{geodesic}) gives the expression for effective potential $V_{\textup{eff}}(r)$, that determines the constants of motion and  the location of the last stable orbit (LSO) \cite{Johannsen:2013szh}:
\begin{align}\label{veff}
\Big(\frac{dr}{d\tau}\Big)^2 \equiv V_{\textup{eff}}(r) = -\frac{1}{2}(g^{tt}E^{2}-2g^{t\phi}EL_z+g^{\phi\phi}L_z^2+1).
\end{align}
We know that eccentric motion exhibits two turning points, namely `periastron' ($r_p = p/(1+e)$) and `apastron' ($r_a = p/(1-e)$) with bounded orbits in the range $r_p<r<r_a$, if $V_{\textup{eff}}(r)<0$ \cite{PhysRevD.103.104045}. These points can be expressed in a more general form, which helps overcome divergences at the turning points when solving Eq. (\ref{geodesic}) for orbital motion by eliminating the proper time ($\tau$) \cite{PhysRevD.50.3816}. We will use the following parametrization of the orbit,
\begin{align}\label{parametrization}
r = \frac{p}{1+e \cos\chi},
\end{align}
where $\chi \in (0, 2\pi)$ helps to determine the orbital period. The points ($\chi=0, \chi=\pi$) correspond to ($r_p, r_a$). Now, using Eq. (\ref{veff}) at the turning points ($r_p, r_a$), we compute the orbital energy and angular momentum. With the constants of motion ($E, L_z$) identified, the LSO of the inspiraling object can be calculated using the following condition: $V'_{\textup{eff}}(r_p)=0$ \cite{PhysRevD.103.104045, PhysRevD.77.124050}. This provides a location for the semi-latus rectum ($p$), which indicates the truncation point of the inspiralling object's trajectory, termed as LSO or `separatrix', denoted as $p_{\textup{sp}}$.

\begin{figure}[h!]\centering
\includegraphics[width=5in, height=2.5in]{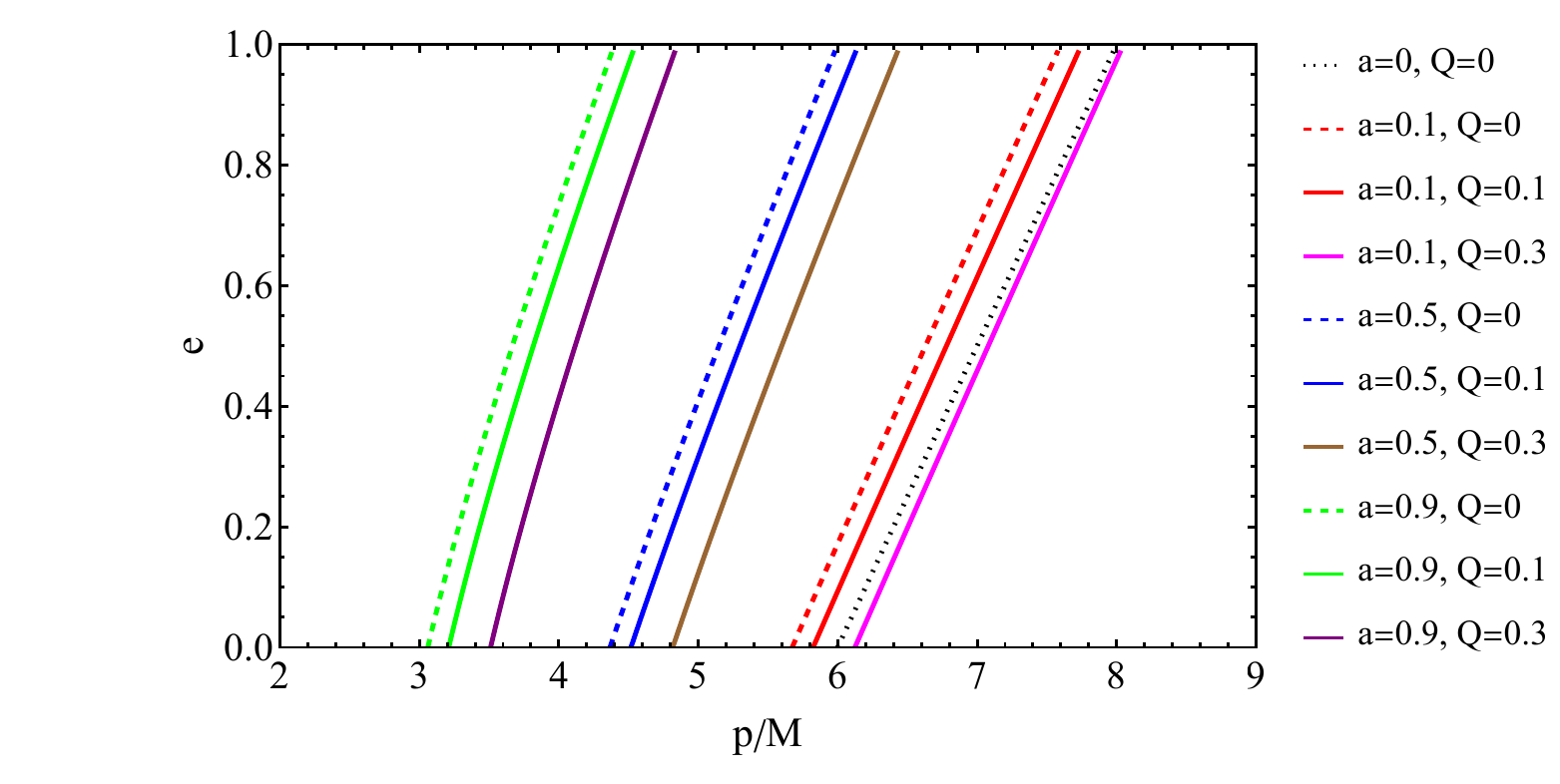}
\caption{The plot shows the separatrix in ($p, e$) plane for distinct values of ($a, Q$). The dotted line with ($a=0, Q=0$) represent the Schwarzschild result $p^{\textup{Sch}}_{\textup{sp}}=6+2e$. However, the dashed lines depict the Kerr results with $Q=0$. The region to the right of the separatrix in the parameter space consists of points that allow bounded orbits, while the separatrix itself defines the boundary of the LSO.} \label{sep}
\end{figure}

We determine ($E, L_z, p_{\textup{sp}}$) numerically since the analytic expressions of the same are extremely lengthy and complicated to mention. However, consistency checks have already been carried out with respect to Schwarzschid and Kerr black holes, which perfectly comply with \cite{Glampedakis:2002ya, PhysRevD.50.3816}. These results will be used in subsequent sections to estimate observables. Moreover, in Fig. (\ref{sep}), we show the effect of black hole spin and tidal charge ($a, Q$) on the separatrix ($p_{\textup{sp}}$) 
It indicates a significant change in the LSO/separatrix as we increase the black hole spin and tidal charge, implicating the noteworthy changes in detectability computation.

Furthermore, we know that eccentric orbits exhibit two fundamental frequencies radial and azimuthal ($\Omega_{r}, \Omega_{\phi}$). This is related to the orbital period of the inspiralling object, i.e. how long it takes for the object to execute the motion from $r_p$ and come back at the same point. The time period can be written as \cite{ PhysRevD.50.3816, Rahman:2023sof}
\begin{align}\label{timeperiod}
T_{P} = \int_{0}^{2\pi}\frac{dt}{d\chi} d\chi, \hspace{1cm} \textup{where,} \hspace{1cm} \frac{dt}{d\chi} = \frac{dt}{dr} \frac{dr}{d\chi},
\end{align}
where $dt/dr$ can be determined by eliminating $\tau$ in Eq. (\ref{geodesic}) and $dr/d\chi$ is obtained from Eq. (\ref{parametrization}). Thus, $T_{P}$ estimates the time period of the object's motion in the background Eq. (\ref{metric}), called the radial time period: $\Omega_{r} = \frac{2\pi}{T_{r}}$. Also, $\Omega_{\phi}\equiv d\phi/dt$ can be obtained from Eq. (\ref{geodesic}). According to these two fundamental frequencies, we can then obtain orbital phase $(\Phi_r,\Phi_\phi)$ with the following equations\cite{PhysRevD.50.3816}
\begin{equation}\label{frequencies}
\frac{d\Phi_{r,\phi}}{dt} = \frac{1}{T_r} \int_0^{2\pi} \Omega_{r,\phi} \frac{dt}{d\chi} d\chi\,.
\end{equation}
Thus, Eq. (\ref{frequencies}) enables us to estimate the orbital phase with the impact of the tidal charge parameter. It is complicated to arrive at the analytical expressions of these quantities, so we use their numerical values in results computed at a later stage of the article.



\section{Teukolsky Perturbation: GW Fluxes and Evolution} \label{perturbation}
As we focus on EMRI systems, the secondary object acts as a perturbation on the spacetime geometry outside the primary, implying that one can use black hole perturbation methods to analyze such setups. In this section, we adopt the Teukolsky perturbation approach to determine observational aspects of measurable quantities through low-frequency GW detectors. Before taking a step further, in general, one can write down the perturbed form of Eq. (\ref{effective EFE}) in the following fashion \cite{Rahman:2022fay},
\begin{align}\label{pert1}
\delta G_{\mu\nu} + \delta E_{\mu\nu} = 8\pi T_{\mu\nu},
\end{align}
where $T_{\mu\nu}$ is given by the energy-momentum tensor of the point particle moving in the background Eq. (\ref{metric}) along the worldline $z^{\mu}$ with proper time $\tau$:
\begin{align}\label{energy-mom}
T^{\mu\nu} = \mu\int\frac{d\tau}{\sqrt{-g}}\frac{dz^{\mu}}{d\tau}\frac{dz^{\nu}}{d\tau}\delta^{(4)}(x-z(\tau)).
\end{align}
Note that there is no energy-momentum tensor involved for the bulk once we consider the vacuum brane. Next, we make the following assumption that the contribution from the bulk is unaltered (or negligible) under the gravitational perturbation on the brane. This statement holds in the low-energy regime, where the matter-energy density on the brane is considered negligible compared to the brane tension, i.e. $\delta E_{\mu\nu}<<\frac{l}{L}E_{\mu\nu}$; where $L$ denotes the curvature length scale of the black hole (primary) on the brane and $l$ is the curvature length scale of the bulk spacetime \cite{Kanno:2003au}. This implies that the bulk geometry is not affected by the gravitational perturbation on the brane, and one can approximate the orbits of the inspiralling object as geodesics. With this meaningful assumption, one can implement the GR Teukolsky framework to construct the perturbation equations.

Next, we provide the separable Teukolsky perturbation equations for the metric Eq. (\ref{metric}). The formalism is preferable for Petrov type D spacetimes where the only non-vanishing Weyl scalar, quantities based on tetrad formalism, is $\Psi_{2}$ and the rest ($\Psi_{0}, \Psi_{1}, \Psi_{3}, \Psi_{4}$) turn out to be zero \cite{Pound:2021qin, PhysRevD.103.124057}, which is also the case for the spacetime under consideration: $\Psi_{2} = \frac{Q-(r+ia\cos\theta)}{(r-ia\cos\theta)^{3} (r+ia\cos\theta)}$; it coincides with Kerr and Schwarzschild cases \cite{Green:2022htq}. The Teukolsky framework is best suited for handling perturbation equations and computing related quantities, as it offers a scalar equation to analyze the dynamics of ($\Psi_{4}, \Psi_{0}$)
 (denoting ingoing and outgoing radiation at the asymptotic) \cite{1973ApJ}. In other words, the formulation reduces the problem to a single perturbation equation than a complex set of tensorial equations. The present article focuses solely on $\Psi_{4}$, i.e. outgoing gravitational waves measured by the detectors at asymptotic infinity. Following \cite{Sasaki:2003xr, Barsanti:2022ana, 1973ApJ, Rahman:2022fay, PhysRevD.102.024041, Zi:2024dpi}, we can obtain separate equations for radial and angular variables for both tensor and scalar perturbations, given by
\begin{align}\label{pert3}
\Delta^{-s}\Big(\Delta^{s+1}\frac{d}{dr}R^{(s)}_{\ell m\omega} \Big)-V^{(s)}_{T}(r)R^{(s)}_{\ell m\omega} = \mathcal{T}^{(s)}_{\ell m\omega}\,,\nonumber
\end{align}
\begin{align}
\Big[\frac{1}{\sin\theta}\frac{d}{d\theta}\Big(\sin\theta\frac{d}{d\theta}\Big)-\Big(\frac{m+s\cos\theta}{\sin\theta}\Big)^{2}-a^{2}\omega^{2}\sin^{2}\theta-2a\omega s\cos\theta+s+2am\omega+\lambda\Big] {}_{s}S_{\ell m} = 0
\end{align}
where $s\in (0, 1, \pm 2)$, potential $V^{(s)}_{T}(r) = -\frac{K^{2}-2is(r-1)K}{\Delta}-4is\omega r+\lambda$ and $K=(r^{2}+a^{2})\omega-am$. $\lambda$ denotes the eigenvalue of the function $_{s}S^{a\omega}_{\ell m}$, termed spin-weighted spheroidal harmonic, which holds the following normalization identity: $\int_{0}^{\pi}\vert _{-2}S_{\ell m}\vert^{2}\sin{\theta}d\theta = 1$. $\mathcal{T}_{\ell m\omega}$ is the source term that we mention in the appendix (\ref{appen}). It is worth mentioning that our analysis completely relies on the gravitational perturbation with outgoing waves; hence, we set $s=-2$ from now and onwards. However, $s=2$ indicates the perturbation equation for $\Psi_{0}$, $s=0$ corresponds to the perturbation for scalar field and $s=1$ describes equations for the electromagnetic field. Since the perturbation equations are well-established in literature and have been used extensively in numerous contexts, readers are suggested to take a look at \cite{Barsanti:2022ana, Sasaki:2003xr, Pound:2021qin} for basic  details.

\begin{figure}[h!]
\centering
\includegraphics[width=3.17in, height=2.2in]{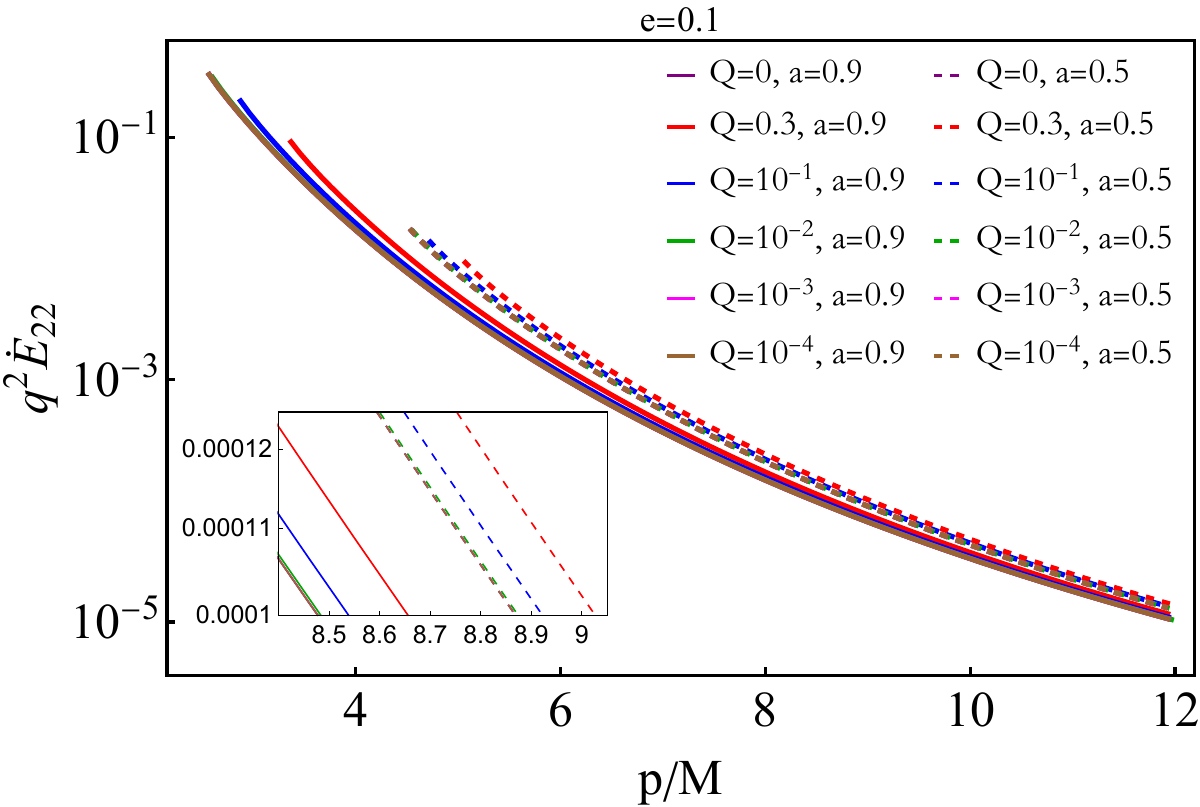}
\includegraphics[width=3.17in, height=2.2in]{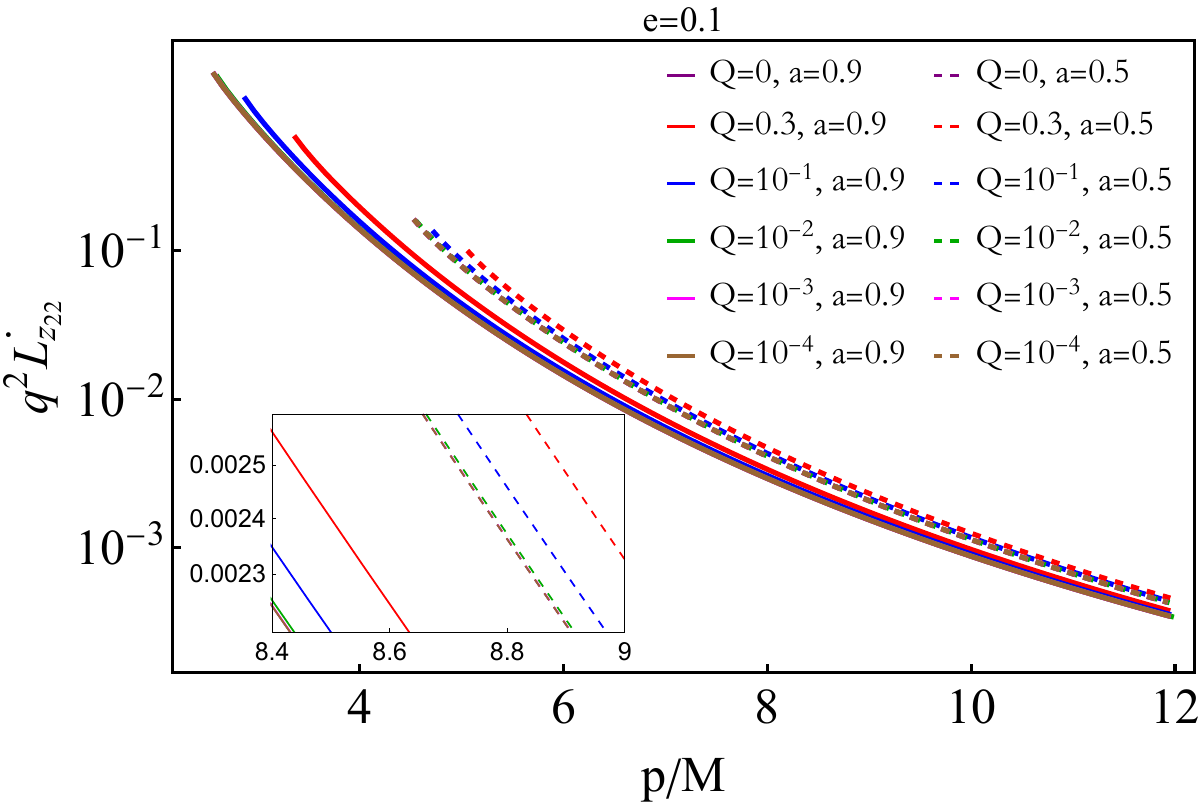}
\includegraphics[width=3.17in, height=2.2in]{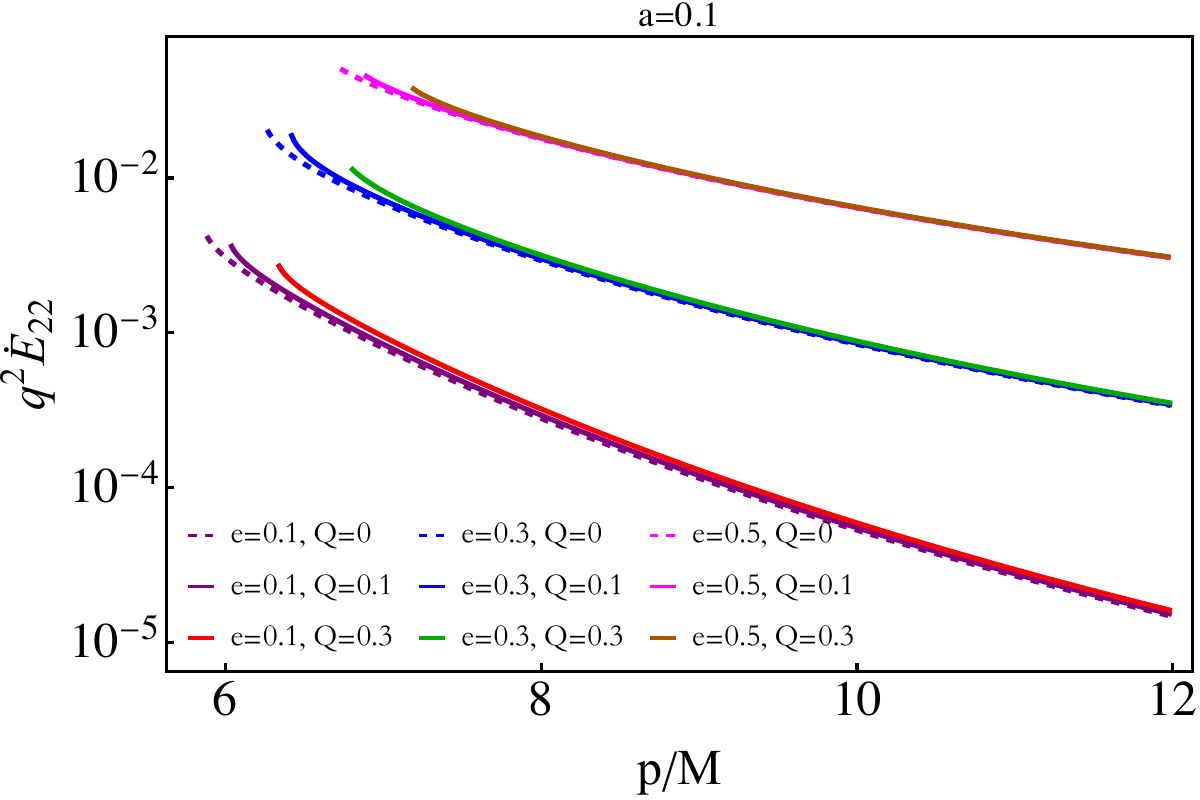}
\includegraphics[width=3.17in, height=2.2in]{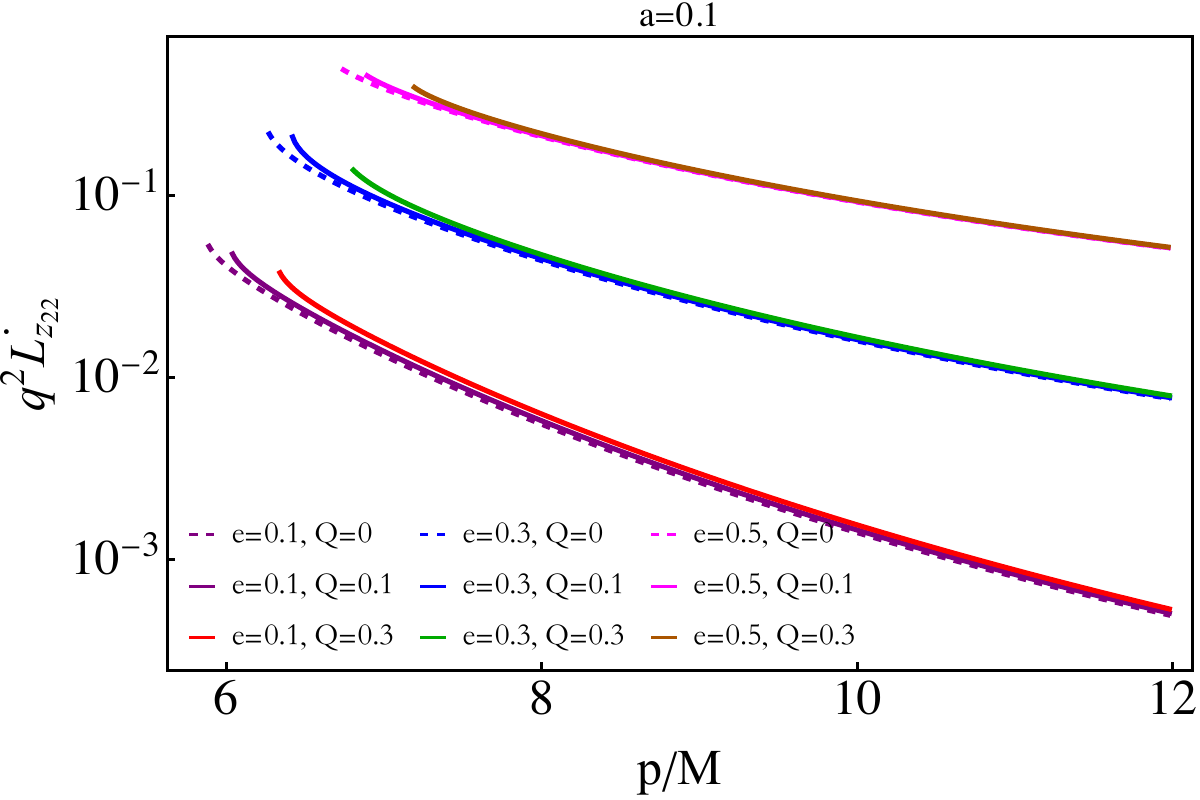}
\caption{Plots show the energy and angular momentum fluxes for the dominant mode (2, 2, 2) for different values of ($e, a, Q$) and their modification from the GR, where the inspiral starts from $p=12$.}\label{Enrflx2}
\end{figure}

The radial equation (\ref{pert3}) admits two homogeneous solutions \cite{Hughes:1999bq, Pani}
\begin{equation}\label{rty1}
  R^{in}_{\ell m\omega}(r)\sim\begin{cases}
    B^{tran}_{\ell m\omega}\Delta^{2}e^{-ik r_{*}}; & \text{$r\longrightarrow  r_{+}$},\\
    B^{out}_{\ell m\omega}r^{3}e^{i\omega r_{*}}+B^{in}_{\ell m\omega}r^{-1}e^{-i\omega r_{*}}; & \text{$r\longrightarrow \infty$},
  \end{cases}
\end{equation}
\begin{equation}\label{try2}
  R^{up}_{\ell m\omega}(r)\sim\begin{cases}
    C^{out}_{\ell m\omega}e^{ik r_{*}}+C^{in}_{\ell m\omega}\Delta^{2}e^{-ikr_{*}}, & \text{$r\longrightarrow r_{+}$},\\
    C^{tran}_{\ell m\omega}r^{3}e^{i\omega r_{*}}; & \text{$r\longrightarrow \infty$},
  \end{cases}
\end{equation}
where $k = \omega - \frac{am}{2r_{+}}$ and $\frac{dr_{*}}{dr} = \frac{r^{2}+a^{2}}{\Delta}$. Finally, the general solution of the radial equation can be written using the Green function method that takes the following form,
\begin{align}\label{gensol}
    R_{\ell m\omega}(r) = \frac{1}{W}\Bigg(R^{up}_{\ell m\omega}\int_{r_{+}}^{r}\frac{R^{in}_{\ell m\omega}\mathcal{T}_{\ell m\omega}}{\Delta^{2}}dr+R^{in}_{\ell m\omega}\int_{r}^{\infty}\frac{R^{up}_{\ell m\omega}\mathcal{T}_{\ell m\omega}}{\Delta^{2}}dr\Bigg)
\end{align}
where the Wronskian $W=\Big(R^{in}_{\ell m\omega}\partial_{r_{*}}R^{up}_{\ell m\omega}-R^{up}_{\ell m\omega}\partial_{r_{*}}R^{in}_{\ell m\omega} \Big)$. It is to add that, $R^{in,up}_{\ell m\omega}$ and $\mathcal{T}_{\ell m\omega}$ are functions of radial coordinate $r$. The asymptotic structure of Eq. (\ref{gensol}) at the infinity and the horizon takes the following form \cite{Hughes:1999bq, Pani}
\begin{align}
R_{\ell m\omega}\xrightarrow{r\rightarrow \infty} Z^{H}_{\ell m\omega}r^{3}e^{i\omega r_{*}} \hspace{0.5cm} ; \hspace{0.5cm} R_{\ell m\omega}\xrightarrow{r\rightarrow r_{+}} Z^{\infty}_{\ell m\omega}\Delta^{2}e^{-ik r_{*}},
\end{align}
where
\begin{align}\label{ampl}
Z^{H}_{\ell m\omega} = \frac{1}{2i\omega B^{in}_{\ell m\omega}} \int_{r_{+}}^{\infty}\frac{R^{in}_{\ell m\omega}}{\Delta^{2}}\mathcal{T}_{\ell m\omega} dr \hspace{0.5cm} ; \hspace{0.5cm} Z^{\infty}_{\ell m\omega} = \frac{B^{tran}_{\ell m\omega}}{2i\omega B^{in}_{\ell m\omega}C^{tran}_{\ell m\omega}}\int_{r_{+}}^{\infty}\frac{R^{up}_{\ell m\omega}}{\Delta^{2}}\mathcal{T}_{\ell m\omega} dr.
\end{align}
The fluxes at asymptotic infinity and the horizon are completely determined by ($Z^{\infty}_{\ell m\omega}, Z^{H}_{\ell m\omega}$). For equatorial orbits, we can write: $Z^{\infty,H}_{\ell m\omega} = \sum_{n=-\infty}^{\infty}Z^{\infty,H}_{\ell mn}\delta(\omega - \omega_{mn})$ with $\omega_{mn} = m\Omega_{\phi}+n\Omega_{r}$ \cite{Barsanti:2022ana, Datta:2024vll}. In order to derive the prefactors of Eq. (\ref{ampl}) and make the potential short range, we use the Sassaki-Nakamura (SN) formalism that is best suited for numerical computation \cite{SASAKI198185, SASAKI198268, 10.1143/PTP.67.1788}. Since these are well-established techniques in the literature, we leave it to the readers to follow the approach outlined in \cite{PhysRevD.102.024041, Sasaki:2003xr, Hughes:1999bq}. Nevertheless, we have provided below only a brief discussion of the key equations that govern the fluxes, ensuring clarity in how these fluxes contribute to the overall dynamical evolution of the system. Thus, following \cite{PhysRevD.102.024041, Sasaki:2003xr, Hughes:1999bq, Datta:2024vll, Barsanti:2022ana}, we can obtain GW fluxes near the horizon and at infinity
\begin{equation}\label{fluxes}
\begin{aligned}
\Big(\frac{dE}{dt}\Big)^{\infty}_{GW}  =& \sum_{\ell mn} \frac{|Z^{H}_{\ell mn}|^2}{4\pi \omega_{\ell mn}^2} \hspace{0.5cm} ; \hspace{0.5cm}
\Big(\frac{dE}{dt}\Big)^{H}_{GW} = \sum_{\ell mn}\alpha_{\ell mn} \frac{|Z^{\infty}_{lmn}|^2}{4\pi \omega_{\ell mn}^2}, \\
\Big(\frac{dL_z}{dt}\Big)^{\infty}_{GW}  =& \sum_{\ell mn} \frac{m |Z^{H}_{\ell mn}|^2}{4\pi \omega_{\ell mn}^3} \hspace{0.5cm} ; \hspace{0.5cm}
\Big(\frac{dL_z}{dt}\Big)^{H}_{GW} = \sum_{\ell mn}\alpha_{\ell mn} \frac{m|Z^{\infty}_{lmn}|^2}{4\pi \omega_{\ell mn}^3}\,.
\end{aligned}
\end{equation}
The total flux can be written as $\dot{\mathcal{C}} = \dot{\mathcal{C}}^{H}+\dot{\mathcal{C}^{\infty}}$ where $\mathcal{C}\in (E, L_z)$. Here `dot' denotes the derivative with respect to the coordinate time $t$, e.g. $\dot{\mathcal{C}}\equiv d\mathcal{C}/dt$. Note that in the process of flux computation, we integrate over $\chi$ after using the parametrization mentioned in Eq. (\ref{parametrization}), enabling us to estimate the fluxes at different orbital radii. Finally, with the use of Eq. (\ref{fluxes}) we compute GW fluxes for different values of ($Q, a, e$) and their deviation from the corresponding  results for Kerr, which can be seen in Fig. (\ref{Enrflx2}). This implies the impact of the extra dimension on the estimated observables.
It is essential to emphasize that we focus only on the dominant mode ($\ell, m, n$ ) = ($2, 2, 2$) as the higher modes makes the computation more costly. Generating data for higher modes, especially on the inspiral grid as discussed in the next subsection (\ref{evolution}), requires a significantly huge amount of time, making large-scale calculations impractical. Prioritizing the dominant mode ensures efficient analysis while preserving key physical insights. Also, in most of the binary black hole systems, the dominant mode is a very prominent and significant contributor to the gravitational radiation emitted during the inspiral phase \cite{OBrien:2019hcj}. However, we leave it for future studies to include higher modes with more efficient techniques.
\subsection{Orbital Evolution}\label{evolution}
In this section, we analyze the impact of the tidal charge parameter on orbital evolution. We study this under `adiabatic approximation,' which implies a significant difference between the two time scales - inspiral and orbital. The approximation considers that the inspiral time scale is much larger than the orbital time scale, which enables us to treat the orbital motion as geodesic. In essence, it implies that the radiated energy and angular momentum fluxes are directly linked to the loss in orbital energy and angular momentum in the following way \cite{Datta:2024vll}:
\begin{align}\label{adiabatic}
\Big(\frac{dC}{dt}\Big)^{\textup{(orbit)}} = -\Big(\frac{dC}{dt}\Big)_{GW}, \hspace{0.2cm} \mathcal{C} \in (E, L_z).
\end{align}
This flux-balance sets up a platform to compute the impact of the tidal charge on the orbital evolution of the eccentric parameters ($p(t), e(t)$). Before starting to compute the adiabatic evolution of EMRI trajectories, we set up a two-dimensional rectangular grid for ($p, e$) with a fixed value of tidal charge and black hole spin to generate the data for energy and angular momentum fluxes. In the $e$ direction, we consider 40 points in the range $e\in (0, 0.6)$, whereas in the $p$ direction, we take 50 points in the following fashion \cite{Datta:2024vll}
\begin{align}
p_{i} = p_{\textup{min}}+4(e^{i\Delta u}-1), \hspace{1cm} 0\leq i \leq 49\,.
\end{align}
This was done to start the inspiral from $p=12$ and make the grid denser so that fluxes are interpolated well between any point between $p_{\textup{min}}\leq p\leq 12$. Also, we set $p_{\textup{min}} = p_{\textup{sp}}+0.02$ and $\Delta u=0.02$. We remind the readers that we have written the expressions in the dimensionless unit. It is important to add that the separatrix region is very sensitive in the sense that all the computed quantities change very rapidly, so to make our analysis meaningful, we extend the inner edge of our grid slightly away from the separatrix, spanning the range $p_{\textup{sp}}\lesssim p \lesssim p_{\textup{sp}}+0.02$. In the context of EMRIs, a secondary object inspirals a primary and generates GWs. This process causes a decay in orbital radius and finally reaches a place beyond which the adiabatic approximation does not hold. This typically occurs near the inner edge of our computational grid. To ensure the accuracy and reliability of our analysis, we set the edge of the grid slightly away from the separatrix with an offset of $0.02$. This choice allows us to concentrate on a region where the adiabatic framework can be used and the error that arises from this minute shift, that is, in a small region $p_{\textup{sp}}$ to $p_{\textup{sp}}+0.02$ is negligible. This careful process ensures that our numerical analysis remains robust while avoiding unnecessary complexities in regions where the adiabatic method becomes less effective.

\begin{figure}[h!]
\centering
\includegraphics[width=3.17in, height=2.2in]{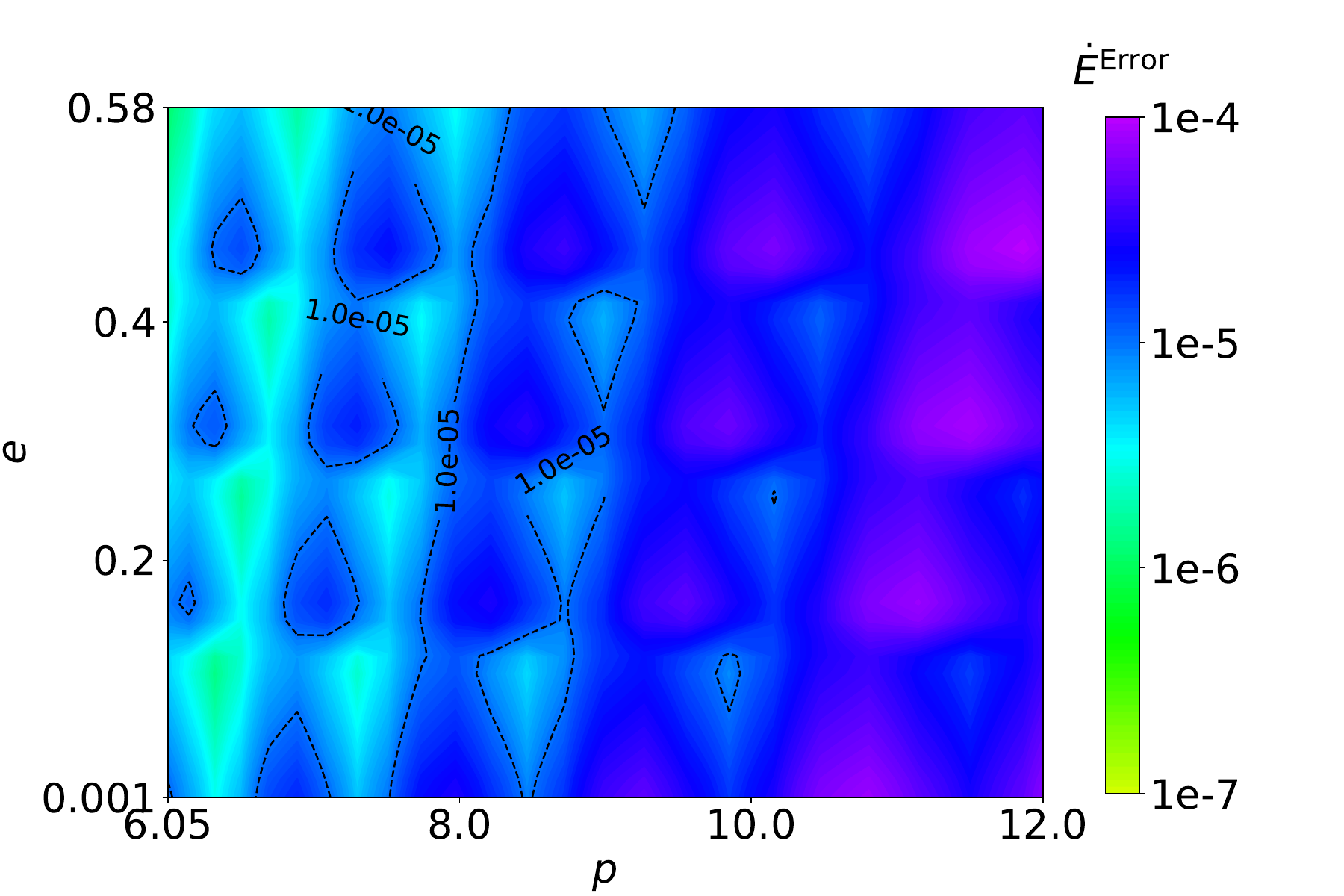}
\includegraphics[width=3.17in, height=2.2in]{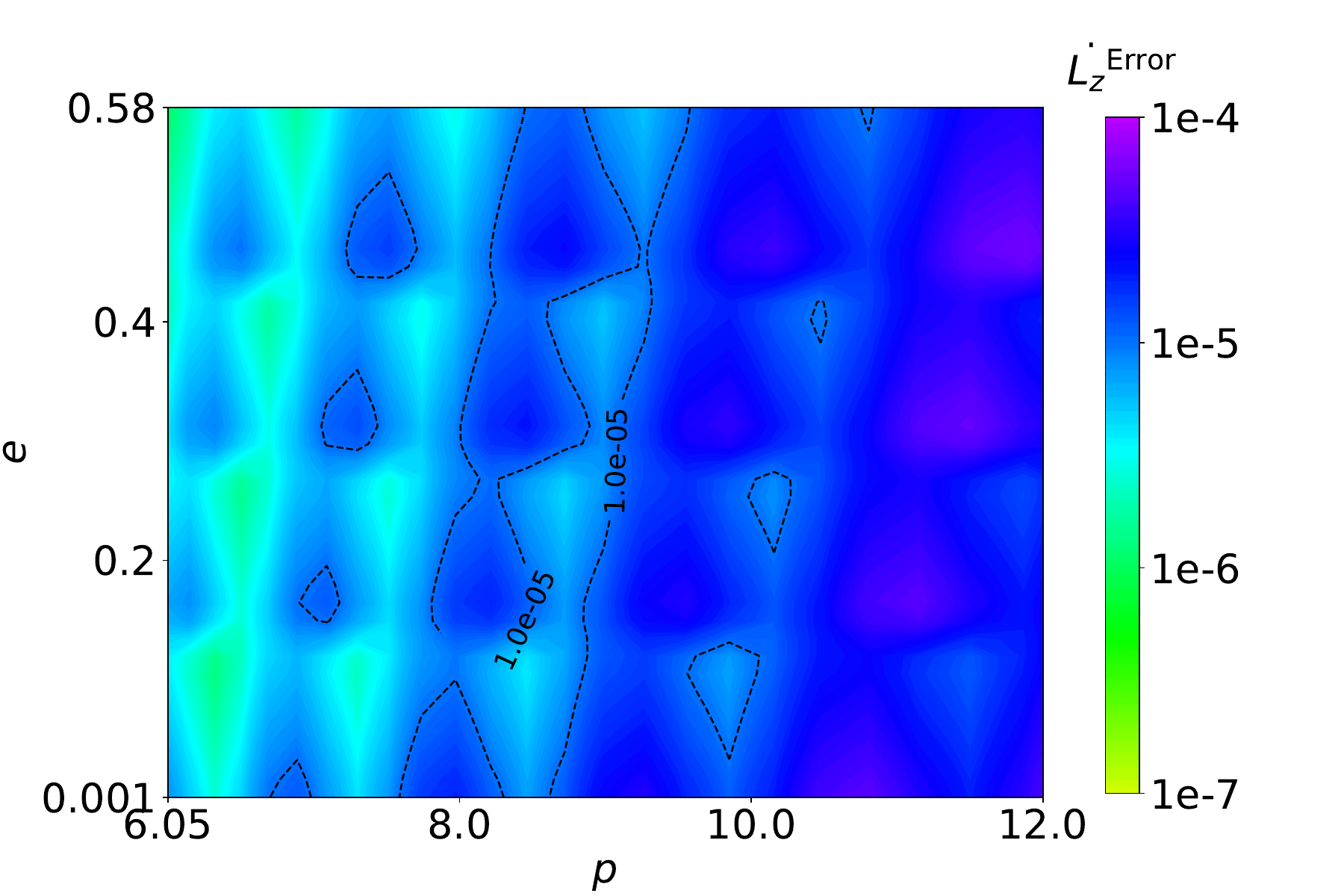}
\caption{The interpolation error of the energy and the angular momentum
fluxes on the sampling grids as a function of orbital parameters $(e,p)$ are plotted, the spin of MBH is $a=0.1$ and tidal charge $Q=10^{-5}$.
} \label{Enrflxerror}
\end{figure}

To carry out the evolution of the orbital parameters, we compute the flux data from Eq. \eqref{fluxes} on the numerical grids, then obtain the flux functions on the plane $(e,p)$ using an interpolation method to generate the fluxes fast. More specifically, assuming the fluxes are approximately smooth functions of orbital parameters in the domain, we construct four interpolants for the energy and angular momentum fluxes near the horizon and at infinity. In Fig. (\ref{Enrflxerror}), we plot the error of interpolation flux functions on the sampling grids for the energy and angular momentum fluxes, where the absolute difference of the flux computed with \eqref{fluxes} and interpolation flux is computed. From the figure, it is evident that the error due to the interpolation is almost less than $10^{-4}$ compared to the flux obtained from Eq. (\ref{fluxes}), so we can claim that the interpolation method provides appropriate results \footnote{We have checked this for various values of $a$ and $Q\,.$}.

\begin{figure}[h!]
\centering
\includegraphics[width=3.17in, height=2.2in]{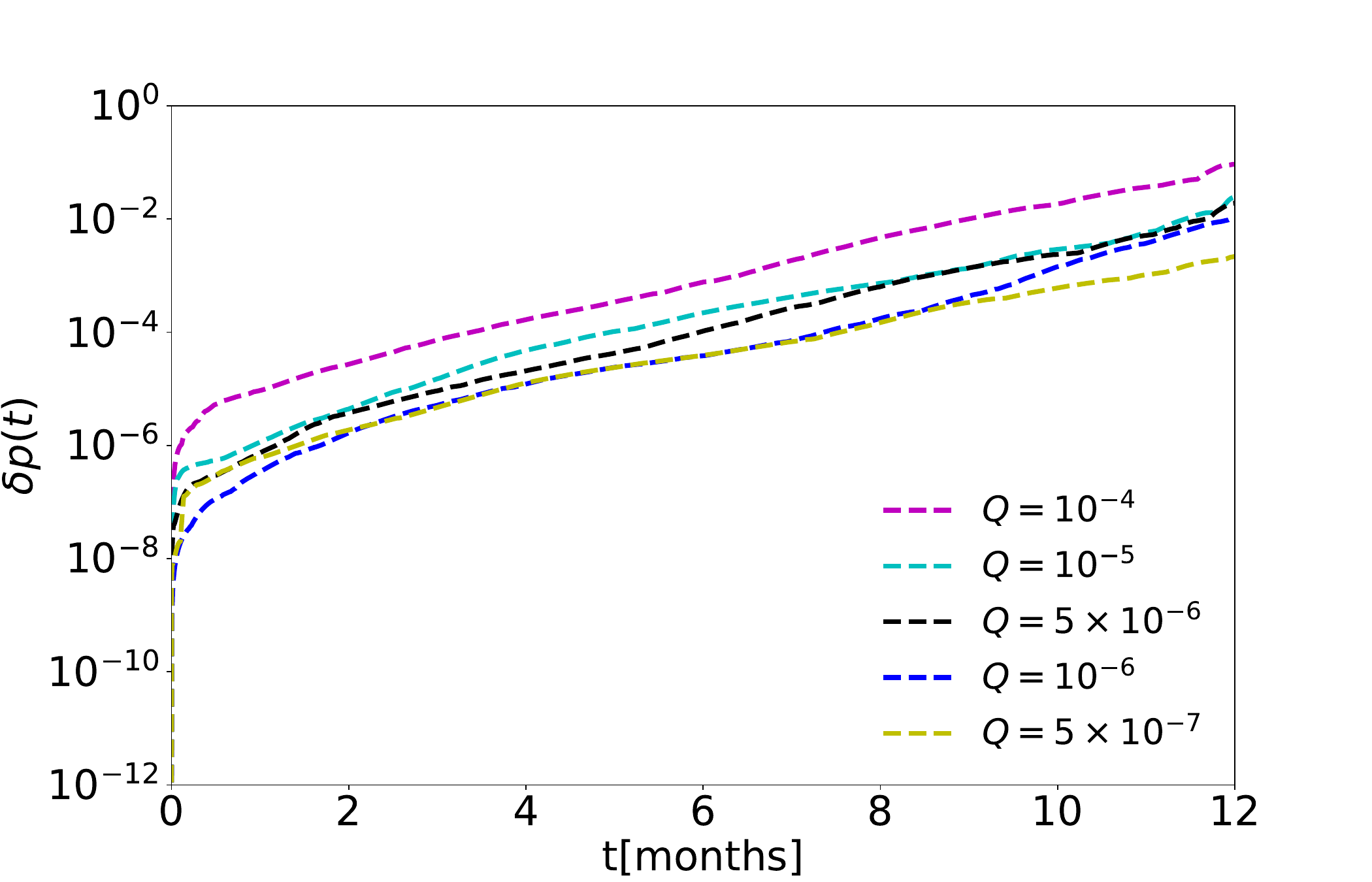}
\includegraphics[width=3.17in, height=2.2in]{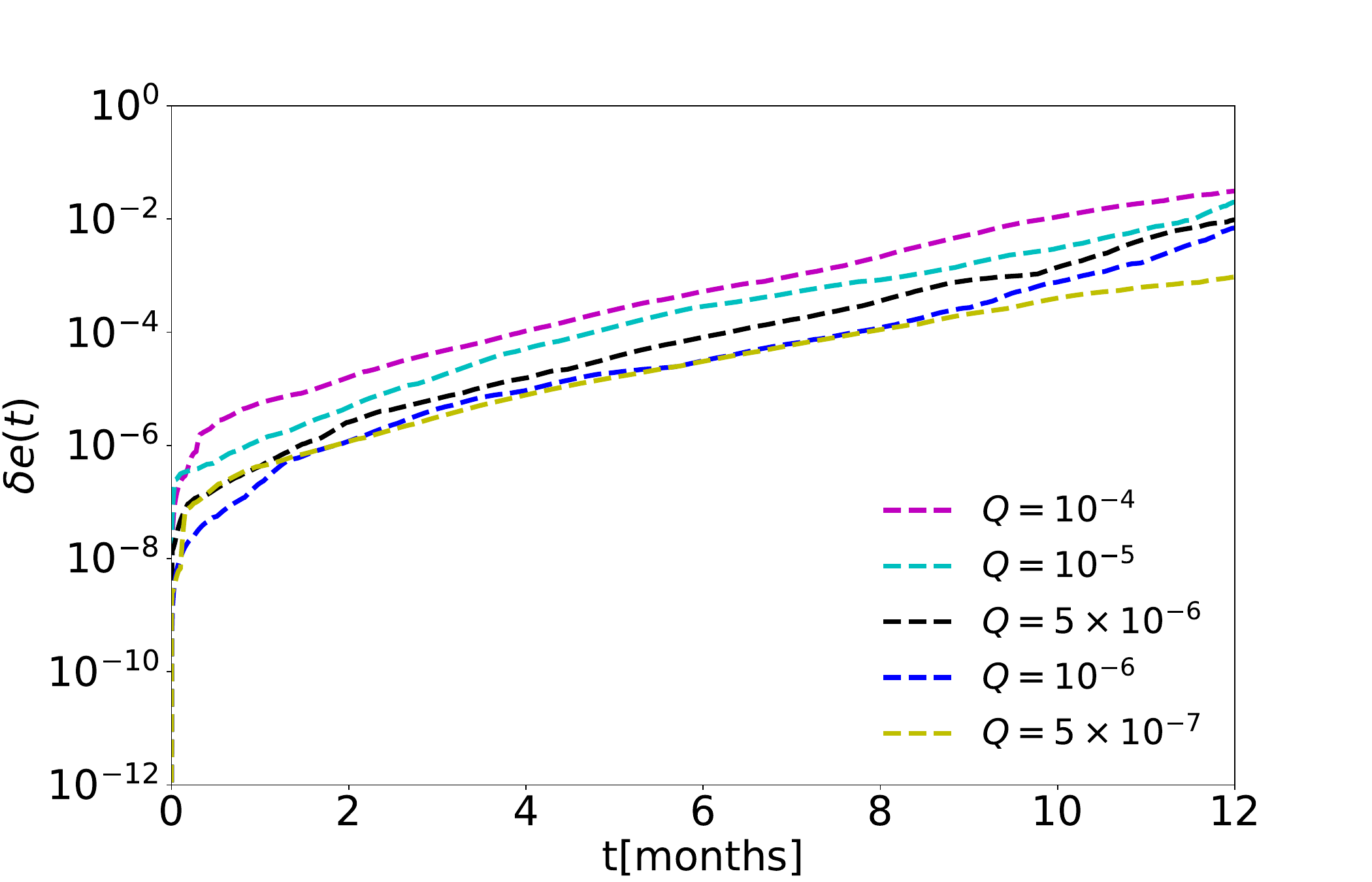}
\caption{The difference of inspiral trajectory due to tidal charge is shown,
the other intrinsic parameters are set as: the spin of the central MBH  $a=0.1$,
the initial orbital semi-latus rectum $p_0=12.0$ and eccentricity, $e_0=0.1$ and tidal charge $Q\in\{10^{-7},10^{-6}, 5\times 10^{-6}, 10^{-6}, 10^{-5}, 10^{-4}\}$.
} \label{traj:diff}
\end{figure}

Now we can construct the evolution of orbital parameters using the following two ordinary differential equations \cite{Hughes:2021exa}
\begin{equation}
\begin{aligned}\label{evol1}
\frac{dp}{dt} &= \frac{(1-e^2)}{2}\frac{dr_a}{dt} +\frac{(1+e^2)}{2}\frac{dr_p}{dt} ,\\
\frac{de}{dt} &= \frac{(1-e^2)}{2p} \Bigg[(1-e) \frac{dr_a}{dt} +(1+e)\frac{dr_p}{dt} \Bigg] ,
\end{aligned}
\end{equation}
where the expressions for $\frac{dr_{a,p}}{dt}$ can be written as
\begin{align}\label{evol2}
\frac{dr_{a,p}}{dt} = \Big(\frac{\partial E}{\partial r_{a,p}}\Big)^{-1}\frac{dE}{dt} + \Big(\frac{\partial L_z}{\partial r_{a,p}}\Big)^{-1}\frac{dL_z}{dt}\,.
\end{align}
Using orbital energy and angular momentum fluxes and Eq. (\ref{evol2}), we estimate the evolution of $e$ and $p$ from Eq. (\ref{evol1}). Hence, using this evolution scheme, we can compute the inspiral trajectories for the orbital parameters $(p(t),e(t))$ and study the impact of the tidal charge on it. To illustrate the effect of tidal charge $Q$ on the orbital dynamics of the EMRIs, we plot the difference between the semi-latus rectum and eccentricity $(\delta p(t), \delta e(t))$ as a function of observation time due to the presence of the tidal charge from the case where it is absent and presented in Fig. (\ref{traj:diff}). One can see that the difference in orbital parameter evolution grows bigger with increasing time, which becomes more obvious for the larger tidal charge. This analysis of trajectory difference helps us to quantitatively assess and understand the significant effect of even small tidal charges.

\section{Detectability: Waveform and Mismatch} \label{wave}
Once orbital evolution is obtained using the method described in the previous section, the corresponding EMRIs waveform can be efficiently computed using the quadrupole formula within the \texttt{FastEMRIWaveforms} (\texttt{FEW}) model \cite{Barack:2003fp,Chua:2020stf,Katz:2021yft}.
For EMRIs signals from the rotating MBH, the \textup{FEW} adopts the \texttt{Augmented Analytical Kludge} (\texttt{AAK}) package \cite{Chua:2017ujo}, allowing for the low-frequency approximate response for the LISA detector with the GPU-accelerated technique.
After all the relativistic adiabatic trajectories are generated within the \texttt{FEW} model, the EMRIs waveform is also accessible. All parameters for EMRI waveform can be summarized as: $$ \lambda_i = (M, \mu, a, p_0, e_0, \Phi_{\phi,0}, \Phi_{r,0}, Q, \phi_S, \theta_S,  \phi_K, \theta_K, d)\,.$$ The angles $(\phi_S, \theta_S)$, $(\phi_K, \theta_K)$ denote the direction of the source position and MBH's spin angular momentum, $d$ is the distance from the source to the detector, and the quantities with subscript `0' represent the initial value. In the \texttt{AAK} model, the waveform is constructed using the quadrupole formula, where the two polarization states in the transverse-traceless gauge are expressed as a sum over $n$-harmonic components \cite{Barack:2003fp}, written as
\begin{align}\label{amplitude}
h_+ \equiv \sum_n A_n^+ = \sum_n &-\Big[1+(\hat{L}\cdot\hat{n})^2\Big]\Big[a_n \cos2\gamma -b_n\sin2\gamma\Big] +c_n\Big[1-(\hat{L}\cdot\hat{n})^2\Big], \nonumber\\
h_\times \equiv \sum_n A_n^\times = \sum_n &2(\hat{L}\cdot\hat{n})\Big[b_n\cos2\gamma+a_n\sin2\gamma\Big]
\end{align}
where $\hat{n}$ is the unit direction vector related to the position of the source and $\hat{L}$ is the orbital angular momentum of the secondary object. The coefficients $(a_n, b_n, c_n)$ are functions of the eccentricity $e$ and mean anomaly $\Phi \equiv \Phi_r$, derived by Peter and Mathews \cite{Peters:1963ux}

\begin{equation}
\begin{aligned}
a_n =~ &-n \mathcal{A} \Big[J_{n-2}(ne)-2eJ_{n-1}(ne)+\frac{2}{n}J_n(ne) +2J_{n+1}(ne) -J_{n+2}(ne)\Big]\cos(n\Phi_r), \\
b_n =~ &-n \mathcal{A}(1-e^2)^{1/2}\Big[J_{n-2}(ne)-2J_{n}(ne)+J_{n+2}(ne)\Big]\times\sin(n\Phi_r), \\
c_n =~& 2\mathcal{A}J_n(ne)\cos(n\Phi_r), \\
\mathcal{A} = ~& (2\pi M \omega_\phi )^{2/3}\mu/d,
\end{aligned}
\end{equation}
where the $J_n$ is Bessel functions of the first kind, $\gamma = \Phi_\phi - \Phi_r$ denotes the direction of eccentric orbit pericenter,
and the quantity $\hat{L}\cdot\hat{n}$ can be given by \cite{Barack:2003fp}
\begin{eqnarray}
\hat{L}\cdot\hat{n} = \cos\theta_S\cos\theta_L + \sin\theta_S \sin\theta_L \cos(\phi_S-\phi_L).
\end{eqnarray}
It should be noted that the angles $(\theta_L, \phi_L)$ are dependent on the parameters $(\theta_K, \phi_K, \Phi_r, \Phi_\phi)$, the full expressions of the same can be found in \cite{Barack:2003fp}. Now, under low-frequency approximation, the EMRI signal responded by LISA can be given by
\begin{equation}
h_{\rm I,II} = \frac{\sqrt{3}}{2} (F^+_{\rm I,II} h^+ + F^\times_{\rm I,II} h^{\times}),
\end{equation}
where the antenna pattern functions $F^{+,\times}_{I,II}$ \cite{Cutler:1994ys} of the detector depend on the orbits of satellites.

Further, to assess the effect of the tidal charge on the EMRIs waveform, we compute the mismatch between the waveform in the presence and absence of the tidal charge parameter. For two waveforms $h_a$ and $h_b$, the mismatch is defined by the inner product in the following way:
\begin{equation} \label{ForMismatch}
\mathcal{M} = 1 - \mathcal{O},
\end{equation}
where $\mathcal{O}$ denotes the overlap between two waveforms, given by
\begin{align}\label{overlap}
\mathcal{O}(h_a, h_b) = \frac{(h_{a}\vert h_{b})}{\sqrt{(h_{a}\vert h_{a})(h_{b}\vert h_{b})}}.
\end{align}
The inner product is defined as \cite{Cutler:1994ys}
\begin{equation}\label{overlap2}
(h_a|h_b) = 2 \int^{f_{high}}_{f_{low}} \frac{h_a^*(f)h_b(f)+h_a(f)h_b^*(f)}{S_n(f)}df\,,
\end{equation}
Eq. (\ref{overlap2}) and related parts are estimated in the frequency ($f$) domain where $f_{\textrm{low}} = 10^{-4} \rm Hz$, $f_{\textrm{high}}$ is the orbital frequency near the LSO ($f_{\rm LSO}$) or the orbital frequency after one year of evolution and $S_n(f)$ is the noise-weighted sensitivity curve of a space-borne detector, such as LISA \cite{LISA:2017pwj}.

\begin{figure}[h!]\centering
\includegraphics[width=6.5in, height=2.5in]{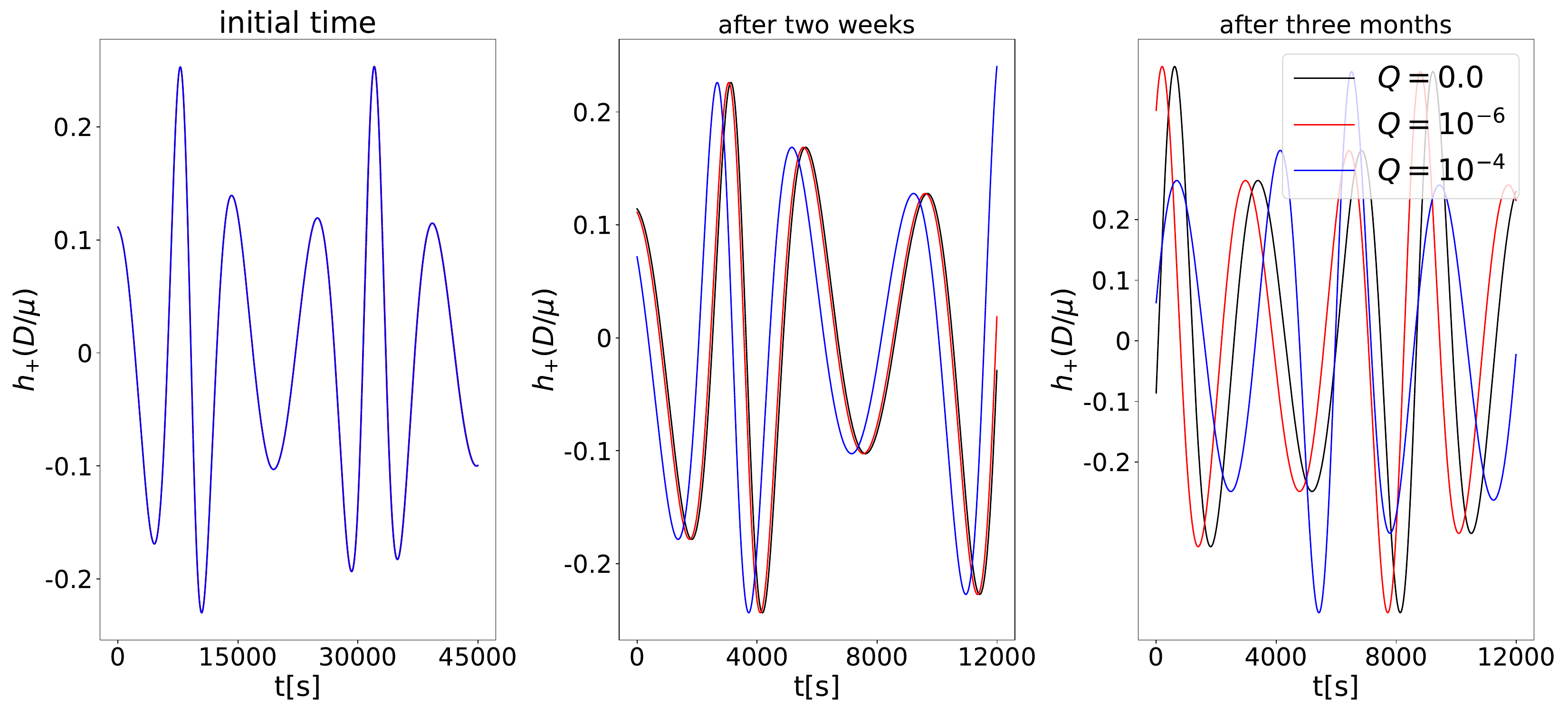}
\caption{The comparison of plus polarization of the EMRI waveform in the time domain for three cases is plotted; the EMRI system has $M=10^6 M_\odot$, $\mu=10 M_\odot$, a=0.1, $p_0=12$, $e_0=0.1$ and $Q \in \{10^{-6}, 10^{-4}\}$.}\label{Figwave}
\end{figure}

In the following section, we set the initial value of orbital phase $\Phi_{r,0}=1.0$, $\Phi_{\phi,0}=2.0$, the mass of the MBH, and the secondary object is $M=10^{6}M_\odot$ and $\mu=10 M_\odot$, respectively. After setting the initial orbital parameters, we generate EMRI waveforms for different cases to compare the effect of tidal charge in Fig. (\ref{Figwave}).
Three waveforms from a rotating MBH with tidal charge $Q\in \{10^{-6}, 10^{-4}\}$ have been plotted for the spin $a=0.1$ and distance $d=1 ~\rm G pc$ with a comparison to the Kerr case.
The figures clearly indicate that the phase exhibits a larger deviation and becomes more noticeable as the evolution proceeds. This effect is amplified for larger values of the tidal charge, where the difference in phase grows more significantly compared to the cases with a smaller tidal charge.

In the next step, we compute the mismatch between two gravitational waveforms, which is crucial in evaluating how well theoretical models represent actual signals. Thus, to evaluate the effect of the tidal charge on the EMRI signal, we show and compare the mismatch between two settings: the first one involves the Kerr MBH without any tidal charge correction, whereas the second captures the effects of the tidal charge. This comparison enables us to quantify the impact of such a parameter on gravitational waveforms together with its potential observational significance. From Eq. (\ref{ForMismatch}), we can infer that when $\mathcal{M} = 0$, two waveforms are identical, implying that the corresponding overlap ($\mathcal{O}$) reaches its maximum value if $\mathcal{O}=1$. However, in real-world scenarios, a small deviation can lead to the introduction of physical effects; for instance, the contribution of extra dimensions (tidal charge). Practically, whether such effects or modifications to GR are detectable by GW detectors or not, a detection threshold is often employed; that is, two waveforms can be distinguished if the mismatch satisfies the condition $\mathcal{M}(h_a, h_b) \leq \mathcal{M}_{\textup{threshold}}\approx \frac{1}{2\rho^{2}}$; where $\rho$ represents the signal-to-noise ratio (SNR), which, for moderate SNR, is $\rho = 20$ \cite{Babak:2017tow,Fan:2020zhy}. This translates into the detection threshold for the mismatch $\mathcal{M}\approx 0.00125$. This sets up a benchmark for examining whether deviations to the GR waveforms, induced by tidal charge, can produce significant differences to be resolved by the detector, thus giving a way to evaluate the observational relevance of such corrections, indicating the existence of extra dimensions

\begin{figure}[h!]
\centering
\includegraphics[width=3.17in, height=2.2in]{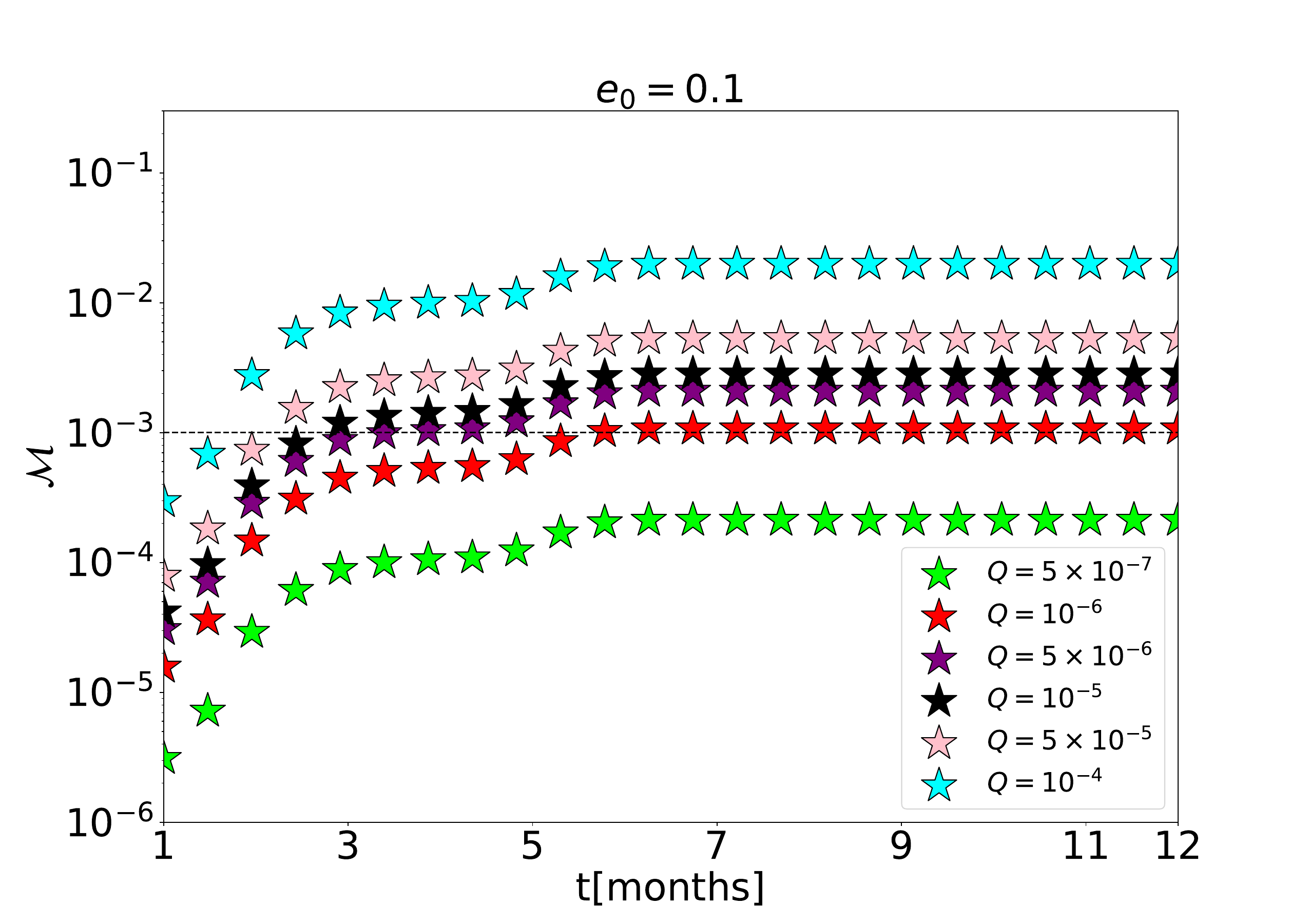}
\includegraphics[width=3.17in, height=2.2in]{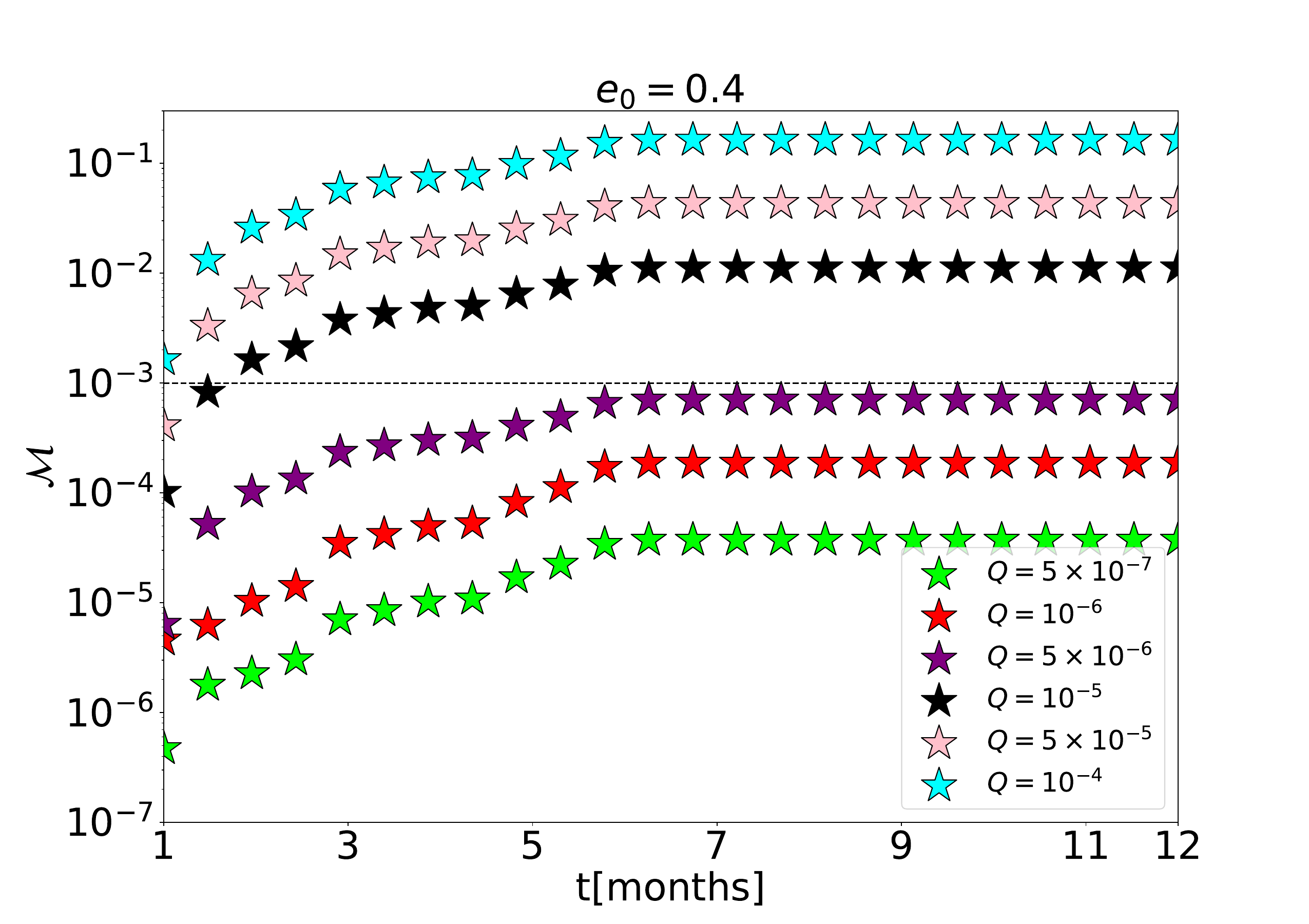}
\caption{Mismatch as a function of observation time for four initial orbital eccentricities $e_0 = (0.1,0.4)$ and MBH's spin $a=0.1$ is plotted, the other parameters are $p_0=12.0$, $Q\in\{5\times 10^{-7}, 10^{-6}, 5\times 10^{-6}, 10^{-5}, 5\times10^{-5}, 10^{-4}\}$.} \label{Fig: mismatch}
\end{figure}

\begin{figure}[h!]
\centering
\includegraphics[width=3.17in, height=2.2in]{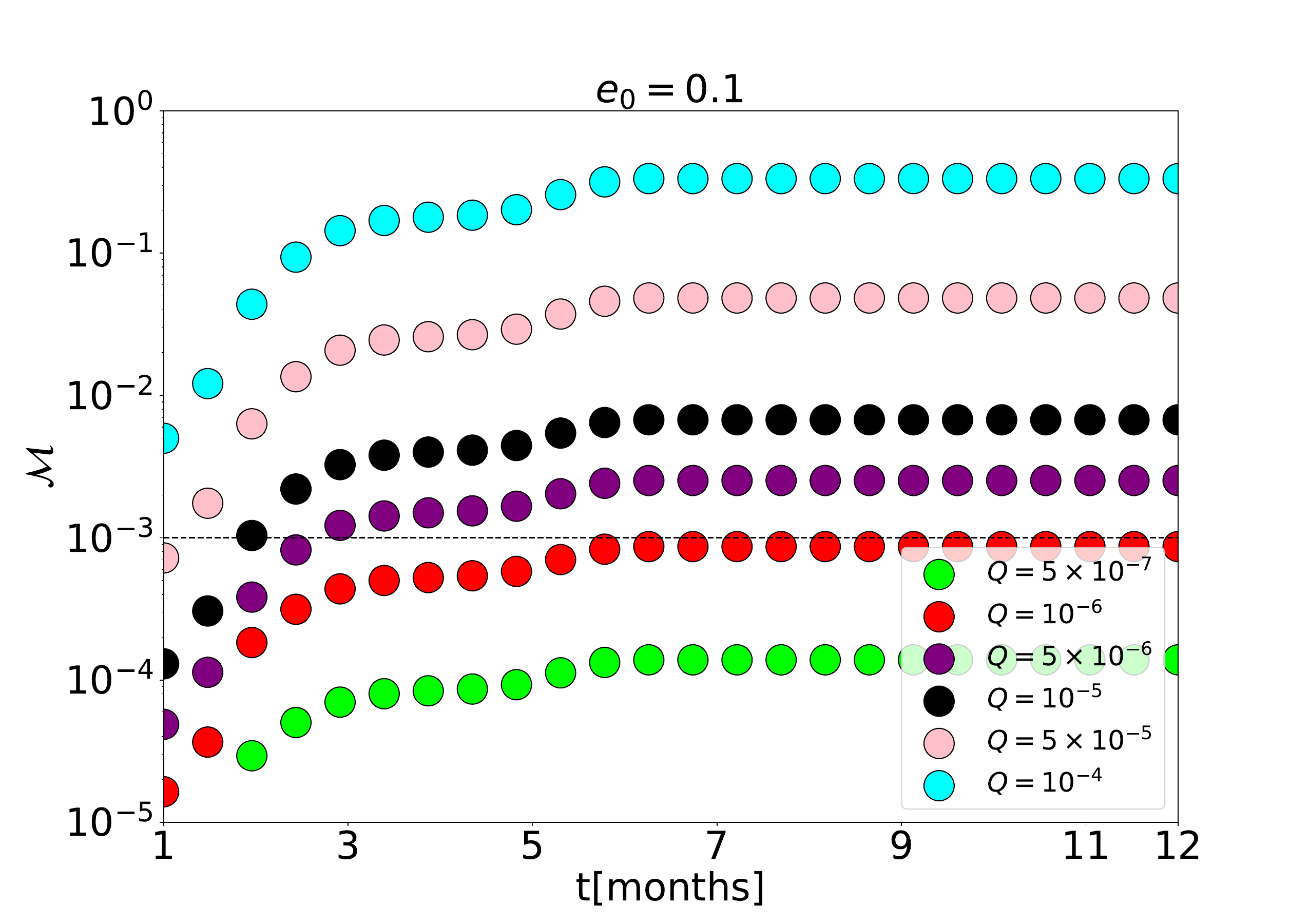}
\includegraphics[width=3.17in, height=2.2in]{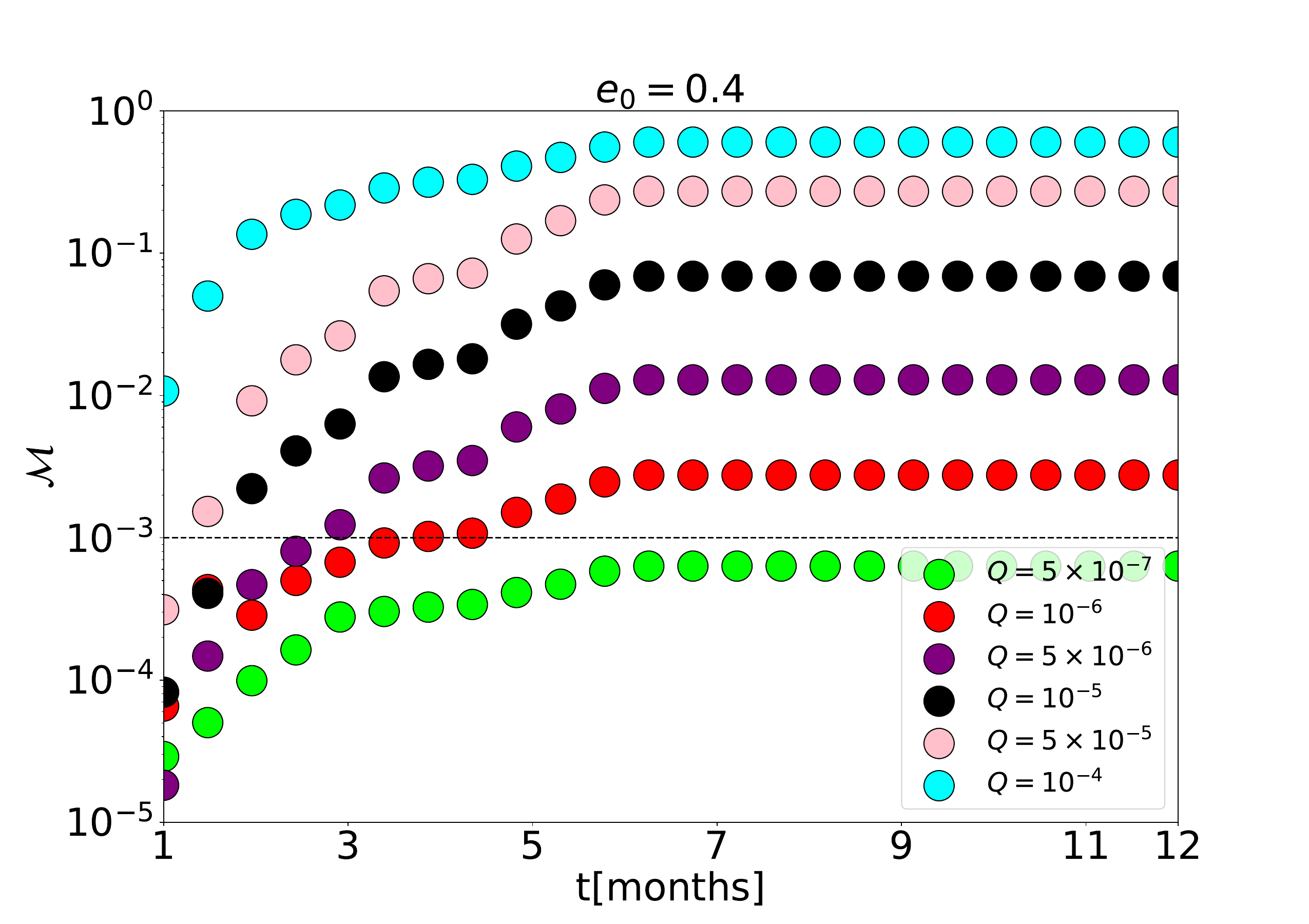}
\caption{Mismatch as a function of observation time  for four initial orbital eccentricities $e_0 = (0.1,0.4)$ and MBH's spin $a=0.9$ is plotted, the other parameters are $p_0=12.0$, $Q\in\{5\times 10^{-7}, 10^{-6}, 5\times 10^{-6}, 10^{-5}, 5\times10^{-5}, 10^{-4}\}$.} \label{Fig: mismatch2}
\end{figure}

As a result, in Fig. (\ref{Fig: mismatch}) and Fig. (\ref{Fig: mismatch2}), we present the mismatch analysis with an observation period of one year and compute the prospects of detecting extra dimensions through GW observations from future low-frequency detectors like LISA. We set the mass-ratio $q=10^{-5}$ ($\mu = 10M_\odot$, $M = 10^{6}M_\odot$) and $p_{0}=12$ as the start of the inspiral. We evaluate the mismatch for small tidal charge values with distinct initial eccentricities, keeping in mind the detectability of the parameter. It is understood from the plots that if we take the larger values of tidal charge, it will lead to a higher mismatch, making the detectability more promising. Note that the horizontal line, representing the detection threshold, allows us to place an upper bound or constraint on the tidal charge in the event that LISA cannot detect the corresponding mismatch. In Fig. (\ref{Fig: mismatch}), fixing $a=0.1$ with $e_{0}=0.1$, LISA can potentially detect the existence of extra dimensions whenever $Q\gtrsim 10^{-6}$, implying an upper bound on the tidal charge in the sense that if we go below the mentioned value, the corresponding mismatch might not be detected. Also, as we increase the initial eccentricity, the estimate on tidal charge increases. For the case with $a=0.1$ and $e = 0.4$, we can see that the order of the tidal charge value effectively turns up $Q\sim 10^{-5}$.

Further, in Fig. (\ref{Fig: mismatch2}),  we present the mismatch plot for higher MBH's spin $a=0.9$ and keep other parameters the same as in the case of Fig. (\ref{Fig: mismatch}). We notice that with ($e=0.1, a=0.9$), we expect an upper bound on the tidal charge $Q\gtrsim 5\times 10^{-6}$ as the corresponding mismatch crosses the detection threshold. Also, in the case of ($e=0.4, a=0.9$), we get the order of magnitude $Q\gtrsim 10^{-6}$, enabling the associated mismatch to cross the threshold value. We have also provided additional plots for different eccentricities and black hole spin in the appendix (\ref{appen2}). Overall, we can notice pronounced changes in the mismatch as we increase eccentricity. This may be because the secondaries easily access the strong field region near the MBH for the more eccentric EMRIs orbit, and the corresponding EMRIs signal can capture richer information about the spacetime; thus, the effect of the tidal charge becomes more evident. For highly spinning MBH, the LSO is closer to MBH, and the EMRIs signal with the correction of the tidal charge has a more obvious deviation from the signal of standard GR. Therefore, our study, which used an order-of-magnitude analysis, including the major impact of eccentricity, indicates the possible detectability of the tidal charge parameter and, consequently, provides a platform for testing theories which predict extra dimensions from LISA observations.

Further, in light of considered values of $Q$, one can perform consistency checks with LIGO-Virgo-KAGRA GWTC-3 observations \cite{LIGOScientific:2021sio}. Taking this into account, given the metric (\ref{metric}) and following the methodology prescribed in \cite{Ryan:1995xi, Flanagan:2007tv, Glampedakis:2002ya, AbhishekChowdhuri:2023gvu}, we find that the leading-order contribution of the tidal charge appears at 1 post-Newtonian (1PN) order. To the leading-order, the average azimuthal frequency takes the following form: $\frac{d\phi}{dt} = \frac{1}{M}\Big(\frac{1-e^{2} }{p}\Big)^{3/2}\Big(1-\frac{Q}{2p}\Big)$. With this, we can easily write down the phase and consequently the dephasing in the frequency domain analytically,
\begin{align}
 \Delta\Phi = \frac{5Q}{96q M\Omega} (-2+(M\Omega)^{2/3}).
\end{align}
Now if we take the mass-ratio in the range of stellar mass black holes, say $q=0.6$, and the central object of $M=140M_\odot$, along with the allowed frequency range (27, 364) Hz mentioned in Table (V) of \cite{LIGOScientific:2021sio}, the corresponding dephasing, with the given frequency range and values of $Q\in (10^{-6}, 0.3)$, goes in the order of $\Delta\Phi\in (10^{-9} - 10^{-4})$. This provides consistency check with the deviation at 1PN order $\delta\phi_{2} \sim 0.05$ as shown in Table (VI) of \cite{LIGOScientific:2021sio}. It is evident that if one takes $Q<10^{-6}$, the obtained dephasing will be $<10^{-9}$. Therefore, in view of the constraint and the meaningful values of $Q$, this justifies that starting from the small values of $Q$ to the larger values $Q\in (10^{-6}, 0.3)$, the dephasing results are consistent with the one obtained from the LIGO data.


\section{Discussion}\label{dscn}
Theoretical advances offer frameworks that allow us to investigate observational consequences and compare studies with actual data from experiments. The future GW observations with space-based low-frequency detectors interestingly aim to investigate the signatures beyond GR, impacting the foundational understanding of gravitational physics. Such non-GR features may imply the presence of additional parameters. In this direction, the concept of braneworld theories appears as one of the riveting directions to examine such notions with extra spatial dimensions. These models can potentially put forward notable modifications to the nature of black hole solutions. In this spirit, a rotating black hole, localized on a 3-brane, provides a prime example to understand the characteristics beyond GR, indicating the existence of extra dimensions. This motivates the importance of studying multidimensional theories, as they offer observational imprints that could help distinguish the existence of extra dimensions from GR. With the recent progress, EMRIs have become a central focus in the search for non-GR effects, attracting considerable interest in GW astronomy. The present article, within the context of EMRIs, analyzes such effects by examining the existence of extra dimensions arising from the braneworld model and their detectability from LISA detectors.

Let us summarize the key findings of the study. With a brief remark on the rotating braneworld metric, we derive the geodesic velocities and fundamental frequencies in terms of orbital parameters. We determine the location of the LSO/separatrix and how it behaves for distinct values of the black hole spin and tidal charge ($a, Q$), as presented in Fig. (\ref{sep}). The separatrix provides us with a region in the ($p, e$) plane to determine the points where we have to truncate the trajectory of the inspiralling object. We implement the Teukolsky perturbation approach to examine the EMRI, where a secondary object ($\mu = 10M_\odot$) exhibits equatorial motion in the background of a massive black hole ($M =10^{6}M_\odot$) which is endowed with an extra-dimensional parameter $Q$. Due to computational challenges involving long runtime for the higher-order modes, our analysis completely focuses on the dominant mode ($2, 2$), which is the strongest component of the GW signal for most of the binaries. With this, we compute the GW fluxes for different values of ($e, a, Q$). Further, within the adiabatic approximation, following \cite{Datta:2024vll, Hughes:2021exa}, we generate fluxes on the ($p, e$) rectangular grid and obtain the impact of the tidal charge on the orbital evolution as presented in Fig. (\ref{traj:diff}). We also examine the interpolation error in the ($\Dot{E}, \Dot{L}_{z}$) on the grids as a function of ($p, e$) in Fig. (\ref{Enrflxerror}). We next employ the evolution data to generate waveforms for different values of the tidal charge, as depicted in Fig. (\ref{Figwave}). We further perform the mismatch analysis to evaluate the possibility of detecting the tidal charge parameter from low-frequency detectors. Considering the detection threshold for the mismatch $\sim 10^{-3}$, we obtain an upper bound with the order of magnitude analysis of the tidal charge. Fig. (\ref{Fig: mismatch}) shows the mismatch for different values of ($e, Q$) with the observation period of one year and mass-ratio $q = 10^{-6}$. We notice that for ($a=0.1, e=0.1$) and ($a=0.1, e=0.4$) the mismatch crosses the detection threshold when $Q\gtrsim 10^{-6}$ and $Q\gtrsim 10^{-5}$, respectively, setting an upper bound on the tidal charge. Further, in the case  ($a=0.9, e=0.1$) and ($a=0.9, e=0.4$), the constraint appears $Q\gtrsim 5\times 10^{-6}$ and $Q\gtrsim 10^{-6}$, respectively as shown in Fig. (\ref{Fig: mismatch2}). It is also apparent from the plots that eccentricity plays a vital role in setting the constraint on the tidal charge. As we increase eccentricity, the corresponding mismatch becomes very significant to be detected, making the study more promising for low-frequency detectors. In fact, we can also observe that the mismatch obtained from the eccentric study is larger than the circular case reported in \cite{Zi:2024dpi, Rahman:2022fay}.

Furthermore, our examination of putting the constraint on extra dimensions is much stronger than the black hole shadow as previously reported in literature \cite{Neves:2020doc}. Although photon orbits in black hole shadows lie close to the event horizon, EMRIs generate long-duration, high-SNR GW signals with thousands of orbital cycles. Their phase evolution, particularly near the separatrix and across a range of radii with eccentric paths, enables wider and more precise spacetime mapping. This makes EMRIs more effective than black hole shadows in constraining non-GR parameters, such as the tidal charge effects considered in our analysis. In \cite{Neves:2020doc}, the most stringent constraint on the tidal charge through shadow analysis has been reported $\lesssim 0.004$.  However, our results through EMRI analysis, with the inclusion of black hole spin and eccentricity of the secondary, deliver a stronger constraint on the tidal charge than the one that exists in literature. Thus, LISA offers a better-suited approach than the shadow and ground-based GW observations for investigating the non-GR aspects and is better equipped to test for the existence of extra dimensions.

Moreover, It is commonly understood that various extensions of GR and beyond can imprint subtle features on EMRIs waveforms; however, the physical mechanisms behind these effects can fundamentally differ. In our work, the deviations arise from extra-dimensional modifications to the metric, contributing to eccentric dynamics through phase shifts, orbital evolution, waveform, and mismatch. In contrast, the recent studies \cite{Collodel:2021jwi, Delgado:2023wnj} focus on the scalar-hairy rotating black holes, where off-center energy densities and scalar configurations reshape the geodesic structure and produce unique features like backward chirps or stalling orbits, effects beyond the scope of the present investigation of our article. These studies obtain and analyze a master equation, which governs the evolution of an EMRIs using the quadrupole hybrid formalism, where a light compact object inspirals toward a central massive object due to GW emission, modeled via the quadrupole formula. Additionally, the study \cite{DellaRocca:2024sda} examines how GWs from EMRIs are affected by a small black hole surrounded by a time-dependent scalar field called a scalar wig. Using a perturbative method for circular orbits around Schwarzschild and Kerr black holes, it is found that the scalar field causes only minor changes to the waveform and detectability by LISA. In the line of these studies comprising a vast motivation of EMRIs, our results suggest that eccentricity amplifies extra-dimensional effects in a way that is distinct from the orbital behavior seen in scalar-hairy scenarios with circular orbits, offering a valuable diagnostic for distinguishing among beyond-GR theories.


Our study opens up several avenues to extend the current framework to a more precise platform where one may efficiently consider higher-order modes to place a more accurate bound on the tidal charge. The ensemble of GW observations, numerical relativity supported with Bayesian statistical methods, and Fisher information matrix (FIM) can be useful tools to perform parameter estimation, correlation among binary parameters, and examine the observational viability of the braneworld model, determining whether signatures arising from extra-dimensional parameters are within the detection range of future observatories. Although such signatures, in general, may appear similar across different models, manifestations of different physical origins. Such overlaps can lead to parameter degeneracies that complicate waveform interpretation and hinder precise model selection. However, these degeneracies can potentially be disentangled when informed by detailed waveform modeling and inspiral data over a long period of evolution, especially including eccentric dynamics. In this spirit, Bayesian framework is crucial for assessing LISA's ability to accurately measure binary parameters. The method, based on the posterior distribution of model parameters, provides a refined understanding of parameters as well as uncertainties that allow us to quantify how well different physical scenarios
 \cite{Thrane:2018qnx, Cornish:2014kda}, such as extra dimensions or scalar fields, align with observed signals. This statistical rigor is essential for disentangling subtle waveform features and isolating signatures of new physics, even in the presence of noise and observational uncertainties. Further, one can consider spinning secondary \cite{Piovano:2020zin} or spin-induced quadrupolar deformation \cite{Rahman:2021eay} to investigate the nature of the secondary object moving in the background of a braneworld black hole. Such analyses can infer the interplay between several parameters starting from the spin of the secondary to ($e, a, Q$), making the analysis more general and wide from the detection perspective, which can reveal subtle effects that become relevant in strong gravity regimes. Furthermore, such an analysis of EMRIs can also be carried out using the Modified Teukolsky framework, which provides a way to improve and strengthen the results more robustly with these advanced techniques. These developments are crucial for space-based detectors like LISA, which are expected to play a vital role in investigating fundamental aspects. We plan to explore some of these studies in future work, as they hold significant promise for advancing our understanding of black hole physics in theories beyond GR.


\section*{Acknowledgements}
The research of S.K. is funded by the National Post-Doctoral Fellowship (N-PDF: PDF/2023/000369) from the ANRF (formerly SERB), Department of Science and Technology (DST), Government of India. T.Z. is funded by the China Postdoctoral Science Foundation with Grant No. 2023M731137 and the National Natural Science Foundation of China with Grant No. 12347140 and No. 12405059. AB is supported by the Core Research Grant (CRG/2023/ 001120) by the Department of Science and Technology Science and Anusandhan National Research Foundation (formerly SERB), Government of India. A.B. also acknowledges the associateship program of the Indian Academy of Science, Bengaluru. S.K. also thanks Rishabh Kumar Singh for useful discussions during the progress of this work.

\appendix

\section{Source Term}\label{appen}
The source term in the Teukolsky perturbation equation (in the background of (\ref{metric})) is written as
\begin{equation}\label{source_term}
\mathcal {T} _ {\ell m \omega} =  4 \int dt d\theta\sin\theta d\phi \frac{\left(B' _ 2 + {B' _ 2}^*\right)}{\bar{\rho}\rho^5} S_{\ell m\omega} e^{-  i (m\phi+ \omega t)}
\end{equation}
where,
\begin{align}
B' _ 2 &= - \frac{1}{2} \rho^8\bar {\rho}\mathcal {L} _{-1}
\bigg[\frac{1}{\rho^4}\mathcal{L}_0\bigg[\frac{T_{nn}}{\rho^2\bar{\rho}} \bigg]\bigg]
 -\frac{1}{2\sqrt{2}}\Delta^2 \rho^8\bar{\rho}\mathcal{L}_ {-1}\bigg[\frac{\bar{\rho}^2}{\rho^4}
J_+\bigg[\frac{T_{\overline{m}n}}{ \hat{\Delta} \rho^2\bar{\rho}^2} \bigg]\bigg] \ , \\
 {B' _ 2}^*&= - \frac {1} {4}\Delta^2 \rho^8\bar{\rho} J_+\bigg[\frac{1}{\rho^4}J_+
\bigg[\frac{\bar{\rho}}{\rho^2}T_{\overline{m}\overline{m}}\bigg] \bigg]
- \frac{1}{2\sqrt {2}}\Delta^2 \rho^8\bar{\rho} J_+ \bigg[\frac{\bar{\rho}^2}{\Delta \rho^4}\mathcal {L}_
{-1}\bigg[\frac{ T_ {\overline{m}n}}{\rho^2\bar {\rho}^2}\bigg] \bigg] \ ,
\end{align}
where we remind that $\Delta = r^{2}-2Mr+a^{2}+QM^{2}$, $K=(r^{2}+a^{2})\omega-am$, $\rho = (r-ia\cos\theta)^{-1}$, $\bar{\rho} = (r+ia\cos\theta)^{-1}$ and operators.
\begin{align}
J_+ = \frac{\partial}{\partial r}+\frac{i K}{\Delta} \hspace{0.1cm} ; \hspace{0.1cm}\mathcal{L} _s = \frac{\partial}{\partial\theta}+\frac{m}{\sin \theta}+s \cot\theta - a\omega\sin\theta \hspace{0.1cm} ; \hspace{0.1mm}
\mathcal{L} _s^\dagger = \frac{\partial}{\partial\theta}-\frac{m}{\sin \theta} + s
\cot\theta + a\omega\sin\theta.
\end{align}
Quantities ($T_{nn}, T_{\bar{m}n}, T_{\bar{m}\bar{m}}, T_{\bar{m}n}$) represent the projections of the stress-energy tensor onto the Newman-Penrose (NP) tetrad basis, given as
\begin{equation}
\begin{aligned}\label{NP}
l^{\mu} =& \frac{1}{\Delta}(r^{2}+a^{2}, \Delta, 0, a), \hspace{0.5cm} n^{\alpha} = \frac{1}{2\rho^{2}}(r^{2}+a^{2}, -\Delta, 0, a)\,, \\
m^{\alpha} =& \frac{1}{\sqrt{2}\bar{\rho}} (ia\sin\theta, 0, 1, i\csc\theta), \hspace{0.5cm} \bar{m}^{\alpha} = \frac{1}{\sqrt{2}\rho}(-ia\sin\theta, 0, 1, -i\csc\theta).
\end{aligned}
\end{equation}
Following \cite{Sasaki:2003xr, PhysRevD.102.024041, Zi:2024dpi}, we can write down the stress-energy tensor of the point particle (mass $\mu$) in the following way,
\begin{align}\label{source1}
T^{\mu\nu} = \frac{\mu}{\Sigma \sin\theta}\Big(\frac{dt}{d\tau}\Big)^{-1}\frac{dz^{\mu}}{d\tau}\frac{dz^{\nu}}{d\tau}\delta(r-r(t))\delta(\theta -\theta(t))\delta(\phi -\phi(t)),
\end{align}
where $z^{\mu} = (t, r(t), \theta(t), \phi(t))$ is the geodesic trajectory with the proper time $\tau = \tau(t)$. Eq. (\ref{source1}) uses geodesic velocities, which can be replaced using Eq. (\ref{geodesic}). Thus the tetrad components of the stress-energy tensor are given by:
\begin{equation}
\begin{aligned}
T_{nn} =& \frac{\mu C_{nn}}{\sin\theta}\delta(r-r(t))\delta(\theta-\theta(t))\delta(\phi-\phi(t)), \\
T_{\bar{m}n} =& \frac{\mu C_{\bar{m}n}}{\sin\theta}\delta(r-r(t))\delta(\theta-\theta(t))\delta(\phi-\phi(t)), \\
T_{\bar{m}\bar{m}} =& \frac{\mu C_{\bar{m}\bar{m}}}{\sin\theta}\delta(r-r(t))\delta(\theta-\theta(t))\delta(\phi-\phi(t)),
\end{aligned}
\end{equation}
where
\begin{equation}
\begin{aligned}
C_{nn} =& \frac{1}{4\Sigma^{3}}\Big(\frac{dt}{d\tau}\Big)^{-1} \Big(E(r^{2}+a^{2})-aL_z+\Sigma\frac{dr}{d\tau} \Big)^{2}, \\
C_{\bar{m}n} =& -\frac{\rho}{2\sqrt{2}\Sigma^{2}}\Big(\frac{dt}{d\tau}\Big)^{-1} \Big(E(r^{2}+a^{2})-aL_z+\Sigma\frac{dr}{d\tau} \Big)\Big(i\sin\theta(aE-L_{z}\csc^{2}\theta)+\Sigma\frac{d\theta}{d\tau}\Big), \\
C_{\bar{m}\bar{m}} =& \frac{\rho^{2}}{2\Sigma}\Big(\frac{dt}{d\tau}\Big)^{-1}\Big(i\sin\theta (aE-L_{z}\csc^{2}\theta)+\Sigma\frac{d\theta}{d\tau} \Big)^{2}.
\end{aligned}
\end{equation}
As mentioned earlier, we use Eq. (\ref{geodesic}) to replace the geodesic velocities. Also, note that we replace the parametrization (\ref{parametrization}) to perform the analysis for eccentric orbits. Thus, it constructs the source term that we use to estimate GW fluxes in section (\ref{perturbation}).

\section{Mismatch}\label{appen2}
\begin{figure}[h!]
\centering
\includegraphics[width=3.17in, height=2.2in]{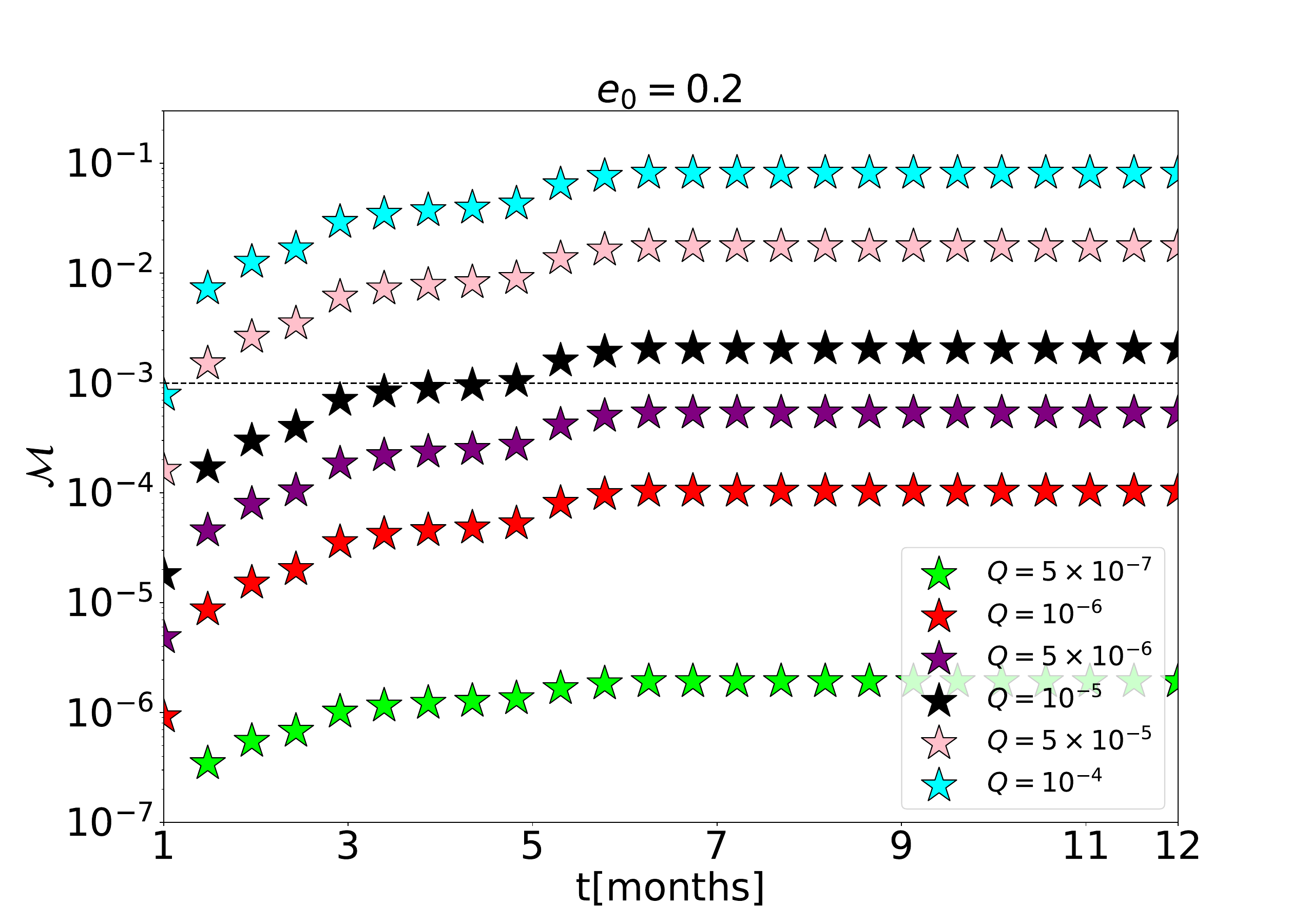}
\includegraphics[width=3.17in, height=2.2in]{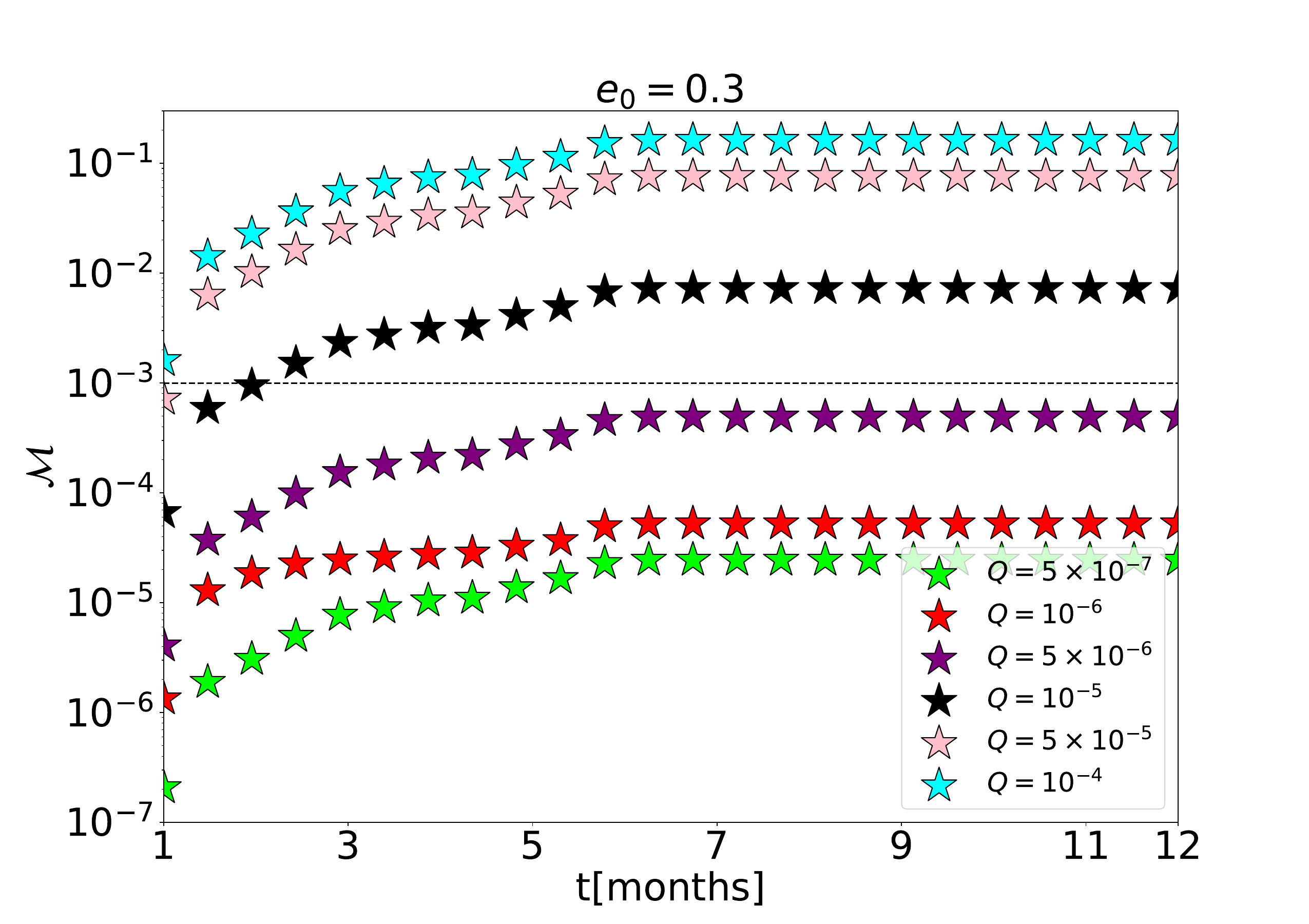}
\caption{Mismatch as a function of observation time  for four initial orbital eccentricities $e_0 = (0.2,0.3)$ and MBH's spin $a=0.1$ is plotted; other parameters are $p_0=12.0$, $Q\in\{5\times 10^{-7}, 10^{-6}, 5\times 10^{-6}, 10^{-5}, 5\times10^{-5}, 10^{-4}\}$.} \label{FigAppen: mismatch}
\end{figure}

Here, we provide additional plots of the mismatch computation for $a=(0.1, 0.9)$ and $e = (0.2, 0.3)$. We can again notice the constraints on the tidal charge, as we increase eccentricity, effectively turns out to be $Q\gtrsim 10^{-5}$ in Fig. (\ref{FigAppen: mismatch}) for $a=0.1$ and $Q\gtrsim 10^{-6}$ in Fig. (\ref{Fig: mismatch2}) for $a=0.9$.

\begin{figure}[h!]
\centering
\includegraphics[width=3.17in, height=2.2in]{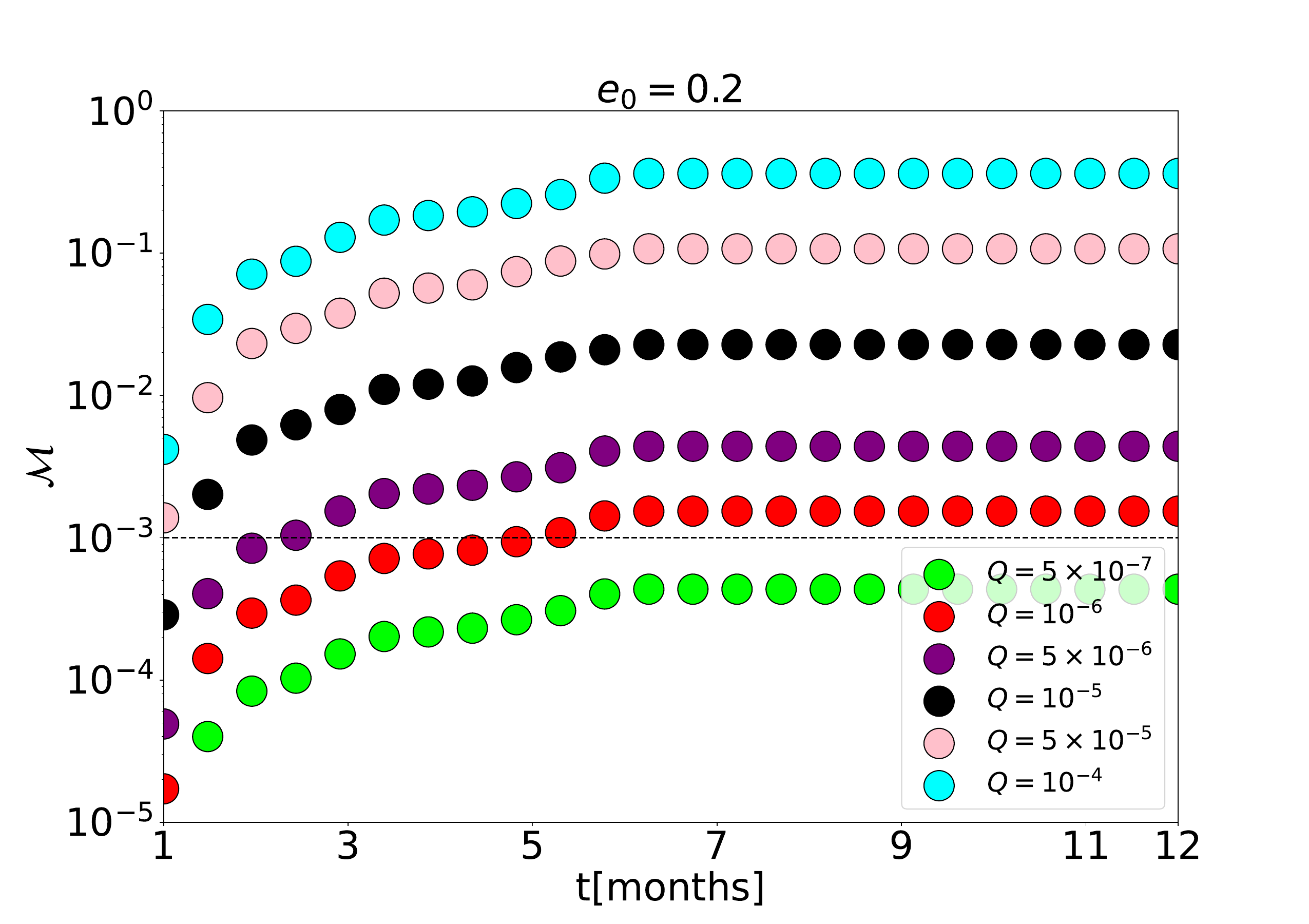}
\includegraphics[width=3.17in, height=2.2in]{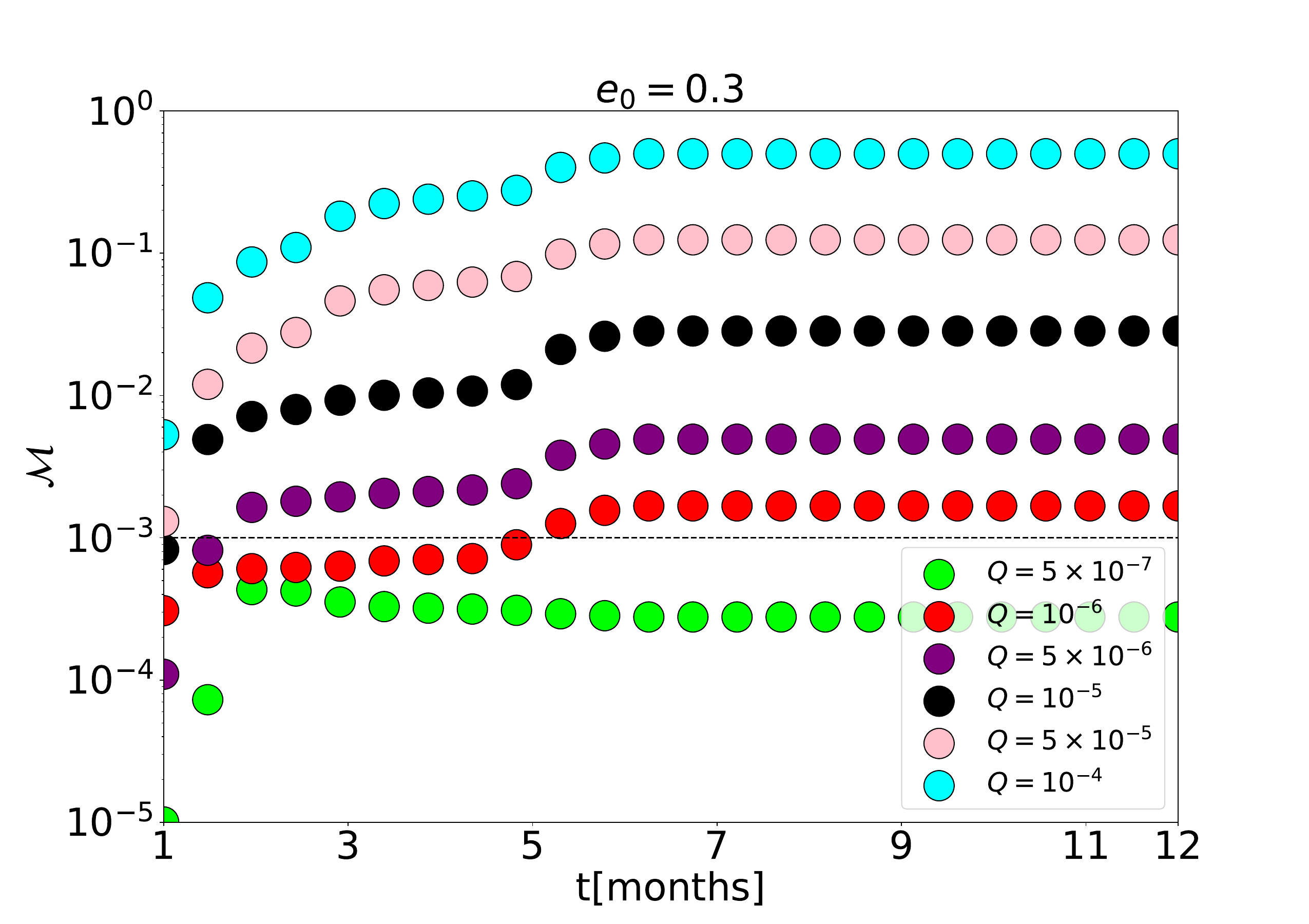}
\caption{Mismatch as a function of observation time  for four initial orbital eccentricities $e_0 = (0.2,0.3)$ and MBH's spin $a=0.9$ is plotted; other parameters are $p_0=12.0$, $Q\in\{5\times 10^{-7}, 10^{-6}, 5\times 10^{-6}, 10^{-5}, 5\times10^{-5}, 10^{-4}\}$.} \label{FigAppen: mismatch2}
\end{figure}
\newpage

\providecommand{\href}[2]{#2}\begingroup\raggedright\endgroup

%

\end{document}